\documentclass[prl,twocolumn,superscriptaddress]{revtex4-1}
\usepackage{amsmath, amsfonts,graphicx,color}

\begin{document}

\newcommand{\todo}[1] {{\color{red} #1}}
\newcommand{\LRC}{$L \leftrightarrow R$}
\newcommand{\I}{\mathcal{I}}
\newcommand{\E}{\mathcal{E}}
\newcommand{\C}{\mathcal{C}}
\newcommand{\OO}{\mathcal{O}}
\newcommand{\R}{\mathcal{R}}
\newcommand{\SC}{\mathcal{S}}	

\title{ Topology, Landscapes, and Biomolecular Energy Transport}
\author{Justin E. Elenewski} 
\affiliation{Biophysics Group, Microsystems and Nanotechnology Division, Physical Measurement Laboratory, National Institute of Standards and Technology, Gaithersburg, MD 20899, USA}
\affiliation{Maryland Nanocenter, University of Maryland, College Park, MD 20742, USA}
\author{Kirill A. Velizhanin}
\affiliation{Theoretical Division, Los Alamos National Laboratory, Los Alamos, NM 87545 USA}
\author{Michael Zwolak}
\email{mpz@nist.gov}
\affiliation{Biophysics Group, Microsystems and Nanotechnology Division, Physical Measurement Laboratory, National Institute of Standards and Technology, Gaithersburg, MD 20899, USA}
\maketitle

\vspace{0.5\baselineskip}
\noindent{\bf Abstract}
\vspace{0.5\baselineskip}

\noindent {\bf While ubiquitous, energy redistribution remains a poorly understood facet of the nonequilibrium thermodynamics of biomolecules. At the molecular level, finite--size effects, pronounced nonlinearities, and ballistic processes produce behavior that diverges from the macroscale. Here, we show that transient thermal transport reflects macromolecular energy landscape architecture through   the topological characteristics of molecular contacts and the nonlinear processes that mediate dynamics. While the former determines transport pathways via pairwise interactions, the latter reflects frustration within the landscape for local conformational rearrangements. Unlike transport through small--molecule systems, such as alkanes, nonlinearity dominates over coherent processes at even quite short time-- and length--scales. Our exhaustive all--atom simulations and novel local--in--time and space analysis, applicable to both theory and experiment, permit  dissection of energy migration in biomolecules. The approach demonstrates that vibrational energy transport can probe otherwise inaccessible aspects of macromolecular dynamics and interactions that underly biological function.}

\vspace{0.5\baselineskip}
\noindent{\bf Introduction}
\vspace{0.5\baselineskip}

\par Biological systems are characterized by a persistent nonequilibrium state, maintained by the open metabolic reactions that drive self--replication.  Directed redistribution of energy is an intrinsic feature, serving to generate  mechanical motion \cite{Andrieux2006, Hwang2017}, mediate allosteric communication \cite{Tu2008,Wang2017,Buchenberg2017}, and drive bioenergetic processes \cite{Ansari1985,Nedergaard2005, Reidel2015}.  The physical scales of these processes can be surprising: Common enzymatic reactions liberate up to 2 eV of heat repeatedly over micro-- to milli--second catalytic cycles \cite{Reidel2015}. This energy is redistributed throughout the surrounding protein scaffold within picoseconds and is either dissipated to mitigate thermally--induced stress, leveraged to induce mechanical motion, or employed to promote further catalytic activity. Irrespective of the endpoint, efficient and directed energy transport is critical to the function of these nanoscale machines.

\par At the macroscale, Fourier's law, $J = -\kappa \nabla T$ and its time--dependent version capture diffusive heat flow, given by the flux $J$, in response to a temperature gradient $\nabla T$. Those two quantities are related by the thermal conductivity $\kappa$ (or the diffusivity $D$), which can be anisotropic. This situation is more complicated at the nanoscale, where competing ballistic and diffusive transport pathways impede a universal description~\cite{Cahill2003, Cahill2013}. In this context, ballistic wavepackets propagate at the speed of sound in a given vibrational band, up the vibrational mean free path, even without the local thermal gradients required for diffusive transport.


\begin{figure*}[t]
\bgroup
\setlength\tabcolsep{8.0pt}
\begin{tabular}{c}
\includegraphics[scale=0.95]{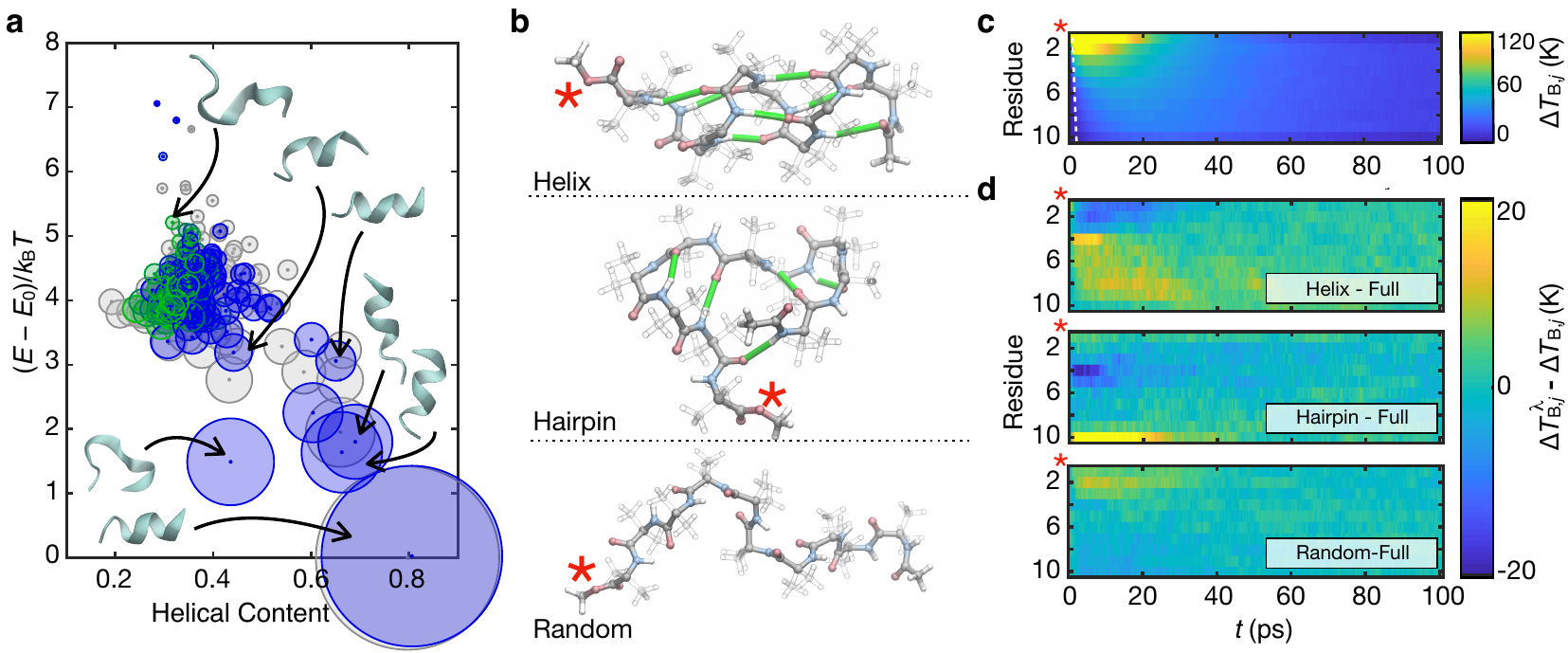} 
\end{tabular}
\egroup
\caption{{\bf Free energy landscapes, topology, and energy transport.} (a) Conformational clusters within the Aib$_{10}$ free energy landscape at the solvent bath temperature $T_\text{B} = 230.0$ K. The size of a data point reflects the relative population of a $k$--means structural cluster at 2.6 nm root--mean--square deviation (RMSD) cutoff.  States for a right--handed helix are colored from blue (more chiral) to green (less chiral), while those of a left--handed helix are uniformly grey.  Helicity parameters and ensemble determinations follow Ref.~\cite{Elenewski2018}; (b) Major conformers in the Aib$_{10}$ structural ensemble. The C--terminal heater residue is denoted by a red asterisk (*), and hydrogen bonds are colored green; (c) Thermal transport profile from NEMD simulations, characterized as a per--residue kinetic temperature elevation $\Delta T_{\text{B},j}(t) = \langle T_j(t)\rangle - T_\text{B}$ with respect to the solvent bath.  The dashed, white line demarcates the ballistic front; (d) Differential heat transport between a full structural ensemble and those ($\Delta T^\lambda_{\text{B},j}$) containing only $\lambda=$ helical, hairpin, or unstructured populations.  Upper and lower limits on the temperature elevation (e.g., on $\Delta T_{\text{B},j}$) provide a cutoff for  all values lying above or below the bound, respectively.   } 
\label{fig:fitComponent}
\end{figure*}

\par Despite the ubiquity of energy redistribution and flow in biomolecular systems, experiments are difficult~\cite{Ansari1985,Srajer1996,Botan2007,Helbing2012,Barends2015,Levantino2015}. In a pioneering work, Botan \emph{et al.}~\cite{Botan2007} developed an approach to monitor real--time heat migration in a polypeptide of 2--aminoisobutyric acid (Aib). The approach employs a photoexcitable azobenzene tag as a heater and backbone carbonyl modes as local vibrational thermometers. The results are complex, suggesting  a `dynamical transition' temperature above which transport is enhanced~\cite{Backus2008, Backus2008b, Schade2009, Backus2009}. Quantum and nonequilibrium molecular dynamics (NEMD) simulations support the presence of a transition in transport properties, and also suggest that a classical description is realistic~\cite{Nguyen2010,Kobus2010,Kobus2011, Goj2011} (unlike for small molecules~\cite{Wang2007,Rubtsova2015,Qasim2019,Rubtsov2019,Liu2019}).  However, both the nature of the transition and  mechanism of transport remain unclear, with theory giving conflicting accounts \cite{Botan2007,Backus2008,Nguyen2010,Kobus2010,Schade2012}.

 In this work, we utilize molecular dynamics simulations and a new space-- and time--local analysis method to explore energy propagation in a paradigmatic  polypeptide.  We find that Fourier behavior captures the bulk of transient energy flow, provided that one accounts for the fact that fluxes and diffusivities are temperature dependent. Departures from a simple realization of Fourier's law happen at large temperature gradients, beyond about 15 K/residue, even though transport is still diffusive. The identification of these regimes is not possible through all-atom molecular dynamics alone~\cite{Nguyen2009,Nguyen2010,Kobus2010,Kobus2011,Goj2011,Brinkman2016,Buchenberg2016} or normal-mode analysis (even when treating anharmonicity as a correction)~\cite{Leitner2001,Yu2003,Yu2003b,Yu2005,Leitner2009,Gnanasekaran2011,Leitner2015}. The former does not unravel the atomic--scale mechanisms of transport and the latter reflects dynamics only at potential energy minima~\cite{Yu2005}.  Within this context, we further demonstrate how the graph--theoretic topology of molecular contacts can define directed pathways for molecular energy redistribution. 

\vspace{0.5\baselineskip}
\noindent{\bf Results}
\vspace{0.5\baselineskip}

\noindent{\textbf{Topology and energy propagation pathways.}}  We initiate our investigations using a series of replica--exchange molecular dynamics (REMD) simulations, as the lack of symmetries, granularity, and high-dimensional free energy landscapes of biomolecules necessitate an exhaustive exploration of conformational space~\cite{Wales1998,Wales2006, Wales2015}.  Our simulation system is a ten--residue Aib helix (Aib$_{10}$) solvated by chloroform, similar to experimental efforts~\cite{Botan2007,Backus2008, Backus2008b, Schade2009, Backus2009}. We previously generated  temperature--dependent free energy landscapes for Aib$_{10}$ at high resolution with replica--exchange simulations~\cite{Elenewski2018}.  From the resulting conformational ensemble, we extract 4000 conformers for each environmental (bath) temperature $T_\text{B}$ according to a Boltzmann distribution. This includes structures from both left-- and right--handed folding funnels, ensuring a uniform distribution of configurations (Fig.~\ref{fig:fitComponent}a,b).  We initiate NEMD simulations in a manner that mimics photoexcitation,  distributing $\approx 1.6$ eV of energy between designated vibrational degrees of freedom in each conformer.  This is achieved by thermostatting the C--terminal residue to a temperature $T' = T_\text{B} + \Delta T$, with $\Delta T =  670$ K, while holding the remainder of the system at $T_\text{B}$.  The simultaneous heating of all vibrational degrees of freedom in the heater residue is well--founded, as it yields thermal transport profiles that are indistinguishable from mode--selective heating~\cite{Botan2007, Nguyen2010}.   This excess energy then propagates freely within the microcanonical ensemble (i.e., without thermostatting).  

The conformational ensemble of Aib$_{10}$ comprises three general structural motifs (Fig.~\ref{fig:fitComponent}b) corresponding to (i) $3_{10}$--/$\alpha$--helical conformers ($\approx45$ \% of ensemble) with hydrogen bonding between residue $j$ and residue $j+3$ or $j+4$, respectively; (ii) hairpin--like configurations, with hydrogen bonds between the first and last residues of Aib$_{10}$ ($\approx15$\%); and (iii) unstructured or extended conformers that have no consistent hydrogen bonding ($\approx40$\%) \cite{Elenewski2018}. We index these subensembles with $\lambda$.  This partition is defined by the underlying free energy landscape, and is thus independent of our thermal transport simulations~\cite{Elenewski2018}.

\par In Fig.~\ref{fig:fitComponent}c,d, we present transport profiles for Aib$_{10}$ versus the ensemble--averaged temperature elevation $\Delta T_{\text{B},j}(t) = \langle T_{j}(t) \rangle - T_\text{B}$ of the $j^\mathrm{th}$ residue, or $\Delta T_{\text{B},j}^{\lambda} - \Delta T_{\text{B},j}$ for subensemble $\lambda$. The full-ensemble profile exhibits a weak thermal front that traverses the peptide within $2$ ps, which is also apparent in the helical ensemble (Fig.~\ref{fig:fitComponent}d). This corresponds to backbone propagation at $v = 1.7$ nm ps$^{-1}$, approaching ballistic transport velocities in biomolecular materials and alkyl chains~\cite{Botan2007, Yue2015,Rubtsova2015,Qasim2019,Rubtsov2019}. While this channel is weak, additional ballistic pathways may exist at lower group velocities in different vibrational bands~\cite{Yue2015,Qasim2016}, though these will inevitably be obscured by more prominent diffusive features. There is also rapid transport with both ballistic and diffusive characteristics across hydrogen-bonded regions, which can be seen in the helical and hairpin conformers (see discussion below).


\begin{figure}[t] 
\includegraphics[width=\columnwidth]{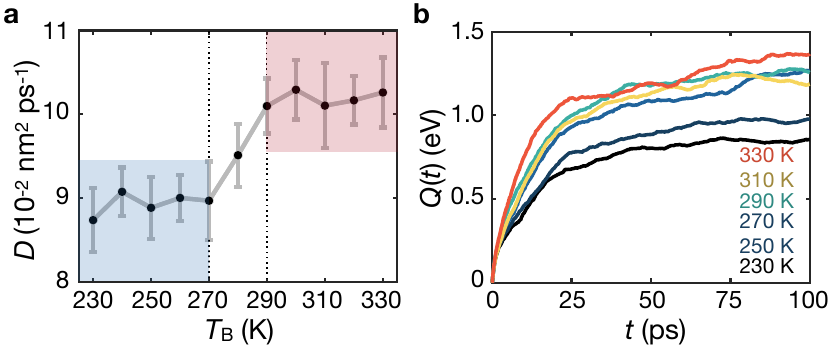} 
\caption{{\bf Benchmarks for thermal transport.} (a) Heat diffusivity $D$ along the major axis of helical Aib$_{10}$ at increasing bath temperatures $T_\text{B}$.  Diffusivities are derived from the time $t_{\text{max}}$ to reach the maximal temperature at each residue following a model $t_{\text{max}} = d^2 / D$, where $d$ is the distance from the heater site.  Colored regions denote low-- (blue) and high--temperature (red) regimes (error bars are plus/minus one standard error).  (b) Net heat $Q(t)$ transferred from residue two to three versus simulation time and bath temperature, following the scheme of Equation~\eqref{mastereq}.  Error bands for the maximal cumulative integration error, as well as net heat transfer between other residues, are in Supplementary Figures 1--9}. \label{fig:diffuse}
\end{figure}

\par While a ballistic pathway exists, the majority of energy transport is nonetheless diffusive --- yielding a broad profile that is sensitive to both temperature and molecular conformation. We separate diffusive and ballistic behavior by coarse--graining in time (into 100 fs bins), averaging away signatures of very fast dynamics, but retain  spatial coarse--graining into individual amino acid residues. We will develop time--dependent quantitative methods to extract diffusivities, free energies, and other characteristics from temperature--based data. However, to facilitate comparison with prior theory and experiment, we initially calculate diffusivities via the time to reach the maximal temperature for each residue. Considering just the helical subensemble for fitting, the temperature-dependent thermal diffusivity $D(T_\text{B})$ has distinct low-- and high--temperature regimes (Fig.~\ref{fig:diffuse}a), which are also reflected in the net heat transfer (Fig.~\ref{fig:diffuse}b). This qualitative behavior agrees with experimental~\cite{Botan2007, Backus2008b,Backus2009} and theoretical~\cite{Botan2007, Nguyen2010,Kobus2010} efforts. These, though, report diffusivities of 0.02~nm$^2$~ps$^{-1}$ and 0.1~nm$^2$~ps$^{-1}$, respectively.  Theoretical $D(T_\text{B})$ from this type of estimate consistently exceed experimental values for Aib$_{10}$ but are comparable to bulk materials~\cite{Rubtsov2019} and other proteins~\cite{Leitner2015}. Force-field parameterization likely contributes to this discrepancy in part. We will see, through an alternate analysis, that residual ballistic components also play a role.   The crossover near 270 K is consistent with prior efforts, which ascribe this behavior to a glass--like dynamical transition~\cite{Botan2007,Backus2008b,Backus2009,Nguyen2010}. We will return to this point.

\par Given this diverse ensemble, it is natural to ask how transport behaves in different conformers. This question was not addressed by prior computational efforts, as they remained below the timescale for structural interconversion in forming their ensemble, sampling only helical configurations and thus a fixed secondary connectivity~\cite{Botan2007,Nguyen2010}. Figure~\ref{fig:fitComponent}d shows the transport profile of the full Aib$_{10}$ ensemble compared to ensembles that contain only helical motifs, hairpin motifs, or randomly oriented conformers without fixed secondary structure. On a residue--by-residue basis, helical conformers propagate heat more readily than the full ensemble.  This is evidenced by less energy retention at the heater site for $t \leq 25 $ ps, commensurate with enhanced transfer to its hydrogen--bonded contacts at early times (mostly site 4 for the helix).  The randomly oriented conformers transport heat less efficiently, underscored by enhanced energy localization at the first three residues for short times and, later, a rate of energy migration that lies slightly below the full ensemble.  We expect a dominant backbone contribution in this case, as longer range contacts are sporadic.   Hairpin configurations are intermediate with enhanced transport to certain hydrogen bond contacts (site 10), in turn reducing the amount of heat transport through others (to the fourth site).  It should be noted that, while hydrogen bonding can lead to more efficient heat transport for certain conformers, backbone channels always carry the majority of heat.  Changes in energy migration are not due to local solvent heating, as the mean temperature of the first two solvation shells increases by at most 5 K over the entire simulation.  While the overall cooling rate involves an interplay between heat diffusivity and surface area--dependent solvent coupling, these effects are minor for the systems considered herein (see the Supplementary Discussion).

\par These observations indicate that topologically nontrivial configurations yield efficient pathways for vibrational energy migration.  The importance of secondary and tertiary contacts has been previously invoked when describing transport within a single conformer of  HP36 \cite{ Leitner2015, Buchenberg2016}.   We extend this observation, demonstrating that representative heat transport characteristics can be obtained only when the conformational landscape is comprehensively sampled.   This is particularly important for metrologies, where insufficient sampling can lead to erroneous diffusivities and the  misidentification of transport pathways. Moreover, changing conditions (temperature, pH, presence of denaturants, etc.)\ can shift the conformational ensemble, particularly near structural transitions. This will be detected by the energy transport, including the capture of additional information about underlying interactions \cite{Velizhanin2011, Chien2013, velizhanin_crossover_2015}. 

\noindent{\textbf{ Heat fluxes and energy landscape topography.}}  While molecular connectivity clearly determines transport pathways, NEMD simulations and existing analysis frameworks afford no immediate means to reconcile temperature--dependent features with microscopic processes and the underlying free energy landscape. To  directly address this, we analyze the intermediate--timescale dynamics of NEMD trajectories -- restricting to helical Aib$_{10}$ conformers for both structural heterogeneity and consistency with prior work -- using a master equation for the kinetic energy $E_j$ of the $j^\mathrm{th}$ residue in the peptide:

\begin{multline} \label{mastereq}
\dot{E}_j(t) =  \sum_{i} [k_{ij}(t) E_i (t)- k_{ji}(t) E_j(t)]  \\ - k_{\text{s},j}(t) [E_{\text{s},j}(t) - E_j(t)].
\end{multline}

\noindent In this case, $k_{ij}(t)$ is a rate constant for energy transfer from residue $i$ to residue $j$ and $k_{ji}(t)$ is a distinct rate for the reverse process (see Methods), $k_{\text{s},j}(t)$ is the rate of  heat transfer to the solvent bath, and $E_{\text{s},j}(t)$ is the kinetic energy density of the solvent surrounding the $j^\mathrm{th}$ residue (scaled to match the residue degrees of freedom).  We diverge from earlier work by treating the $k_{ij}(t)$ as parameters that depend on both position and time --- thereby  implying a temperature dependence.  This accommodation is key to our subsequent analysis.  Given this arrangement, one can identify two distinct intra--peptide couplings: (i) direct transfer between nearest--neighbors in the peptide backbone $(k_{j,j+1}$ and $k_{j,j-1})$ and (ii) a long distance coupling between hydrogen bonding partners ($k_{j,j+3}$, $k_{j,j+4}$ for ideal $3_{10}$-- and $\alpha$--helices, respectively).  With additional approximations, the system in Equation (\ref{mastereq}) becomes well--posed and solvable at all times (see Methods).  This diverges from existing master equation analyses which assume rate constants that are time-- and space--independent, and thus independent of the local temperatures and gradients  \cite{Buchenberg2016}.  These prior works nonetheless treat a broad network of nonlocal contacts, which combined with the analysis here would constitute a logical extension of our methods.

Our remaining discussion is driven by the pairwise heat fluxes $J_{i,j}(t_n) = -k_{i,j}(t_n) [E_i(t_n) - (f_j / f_i) E_j(t_n)]$ and rate constants between coupled residues. Here $f_j$ is the number of degrees of freedom for residue $j$ and  $t_n$ indexes the time domain coarse--graining of the simulation trajectory into $n \leq N$ bins via block averaging.  This approach is a finite difference decomposition of the diffusion equation $\dot{E}(x,t) = D\,\nabla^2 E(x,t)$ at the timescale $\Delta t = t_{n+1} - t_n$ and a length-scale $\Delta x$ defined by the distance between adjacent residues. The fluxes come from the finite difference decomposition of $J(x,t) = -D \, \nabla E(x,t)$. 

The rate constants $k_{i,j}(t_n) = D(t_n) / (\Delta x)^2$, in particular, capture biomolecular heat diffusivity $D(t_n)$ while giving a metric for energy landscape features. We are interested in the distribution of barriers  between low--lying conformational minima, specifically those connected by the energy--transmitting structural displacements that are associated with vibrational energy propagation.  This latter property is reflected by the local, activated conformational changes underlying transport $k_{i,j} = \Omega_{i,j} \,\exp[-\Delta G_{i,j} / k_\text{B} T]$, where $\Delta G_{i,j}$ is the free energy barrier between heat--accepting microstates and $(\Omega_{i,j})^{-1}$ is an effective timescale for free diffusion, influenced by both the protein and its environmental coupling.  While each pair of microstates is characterized by a distinct $\Delta G_{i,j}$, these values evolve during heat transport --- commensurate with changes in the free energy landscape.


\begin{figure}[t] 
\includegraphics[width=\columnwidth]{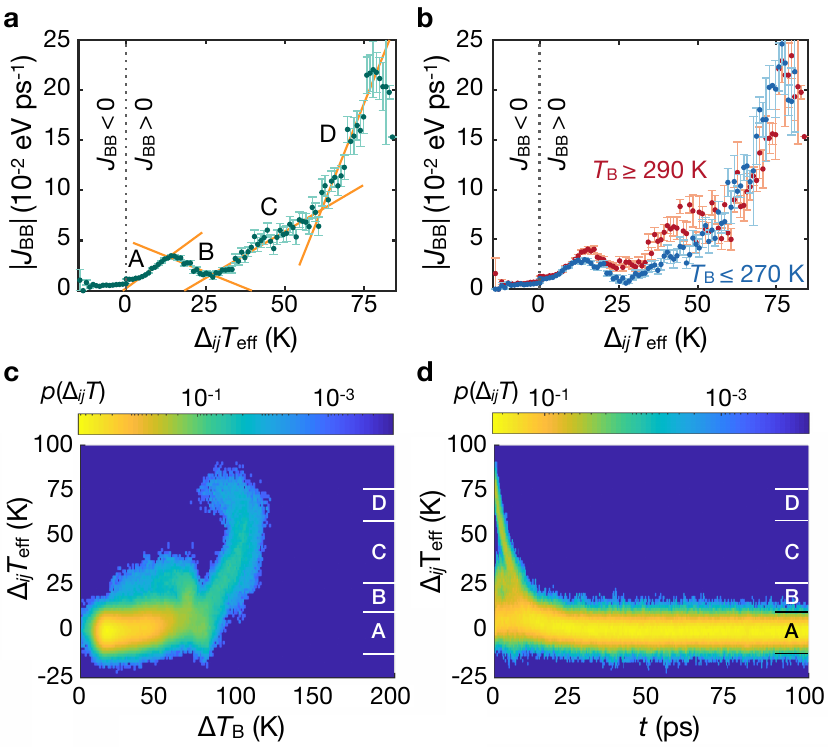} 
\caption{{\bf Flux and thermal gradient distributions.} (a) Backbone flux distributions ($J_\text{BB}$) for helical Aib$_{10}$ conformers.   Fluxes are parameterized by the effective temperature gradient $\Delta_{ij} T_\text{eff}$ between adjacent residues, and a positive flux corresponds to flow away from the heater along the backbone.  Transport regimes are labeled  parallel to the text (A through D) and with orange lines for visibility.  (b)  The $J_\text{BB}$ distribution may be partitioned into low--temperature (blue; 230 K to 270 K) and high--temperature (red; 290 K to 330 K) regimes.     (c) Distribution of local temperature gradients $\Delta_{ij} T_\text{eff}$ versus average elevation $\Delta T_{\text{B},j}(t) = \langle T_j(t)\rangle - T_\text{B}$ over the bath temperature and (d) versus simulation time $t$ for the ensemble of MD trajectories.  Labels (A--D) correspond to the regimes described in the text. Time series data from MD simulation is averaged with $\Delta t = 100$ fs for fits to the  master equation, {Equation~\eqref{mastereq}}, and the resulting fluxes are block averaged in 1.0 K bins. The error bars are plus/minus one standard error.} \label{fig:fluxprofile}
\end{figure}

We employ this kinetic approach with an intermediate timescale ($\Delta t = 100$ fs), long enough to average over most coherent motion but short enough not to obscure the evolution of energy in time. The distribution of backbone fluxes $J_\text{BB}$ is parameterized by an effective temperature gradient $\Delta_{ij} T_\text{eff} = 2[E_i - (f_j / f_i) E_j] / 3N k_\text{B}$  between  residues $i$ and $j$, where the flux is incident on a residue containing $N$ atoms.  While transport is explicitly quantified through $J_\text{BB}$ for simplicity, the effect of hydrogen bonding is present when fitting the backbone flux distribution at hydrogen bonding sites. The results for $J_\text{BB}$ are presented in Fig.~\ref{fig:fluxprofile}a.  A complimentary analysis for $J_\text{HB}$ and a validation of fitting methods are presented in Supplementary Figures 10--15.

{\bf Region A.} The forward flux $J_\text{BB}$ has a linear region for small $\Delta_{ij} T_\text{eff}$ (less than about 15 K), although it does not go to zero at $\Delta_{ij} T_\text{eff}=0$.  Purely diffusive transport will not afford a heat flux in the absence of a local temperature gradient.  Thus, a finite $J_\text{BB}$ at $\Delta_{ij} T_\text{eff} = 0$ is a signature of ballistic/coherent behavior.  Supporting this interpretation, we find that the zero--gradient flux to decrease with increasing $\Delta t$ during coarse--graining, while only exhibiting small error bands at all  scales (thus it is not due to short--timescale fluctuations).  A linear fit to this regime gives an effective diffusivity of  $D_{\text{eff},\text{A}} = 2.3 \times 10^{-2}$~nm$^2$~ps$^{-1}$ (or conductivity $\kappa_{\text{eff},\text{A}} = 3.9 \times 10^{-3}$~eV~K$^{-1}$~ps$^{-1}$). Fitting for small $\Delta_{ij} T_\text{eff}$, while ignoring the residual ballistic contribution right around $\Delta_{ij} T_\text{eff}=0$, removes high rate constant artifacts. Encouragingly, the magnitude of the resulting diffusivity is consistent with experimental values \cite{Botan2007,Backus2008}.  Employing the time to reach the maximum temperature, as done in prior theoretical work (see discussion above), affords much higher diffusivities. This linear regime has the same slope regardless of whether the lattice is in the low-- or high--temperature regime (Fig.~\ref{fig:fluxprofile}b). 

The lack of a dependence on temperature indicates that this regime of transport occurs in a lightly corrugated landscape --- that is, with low--lying barriers separating the minima associated with thermal transport.  In this case, the characteristic barrier scale is below 15 meV,  and thus the mean energy at the lowest background temperature ($T_\text{B} = 230$ K) is above the landscape corrugation. Lower temperature observations are necessary to identify the precise scale, requiring an accurate treatment of quantum effects and different experimental protocols. Stated more succinctly, the equality of the low-- and high--temperature diffusivity indicates that the characteristic time $\Omega^{-1}$ is the same and no free energy barrier exists at this level of landscape hierarchy.


\begin{figure}[t] 
\includegraphics[width=\columnwidth]{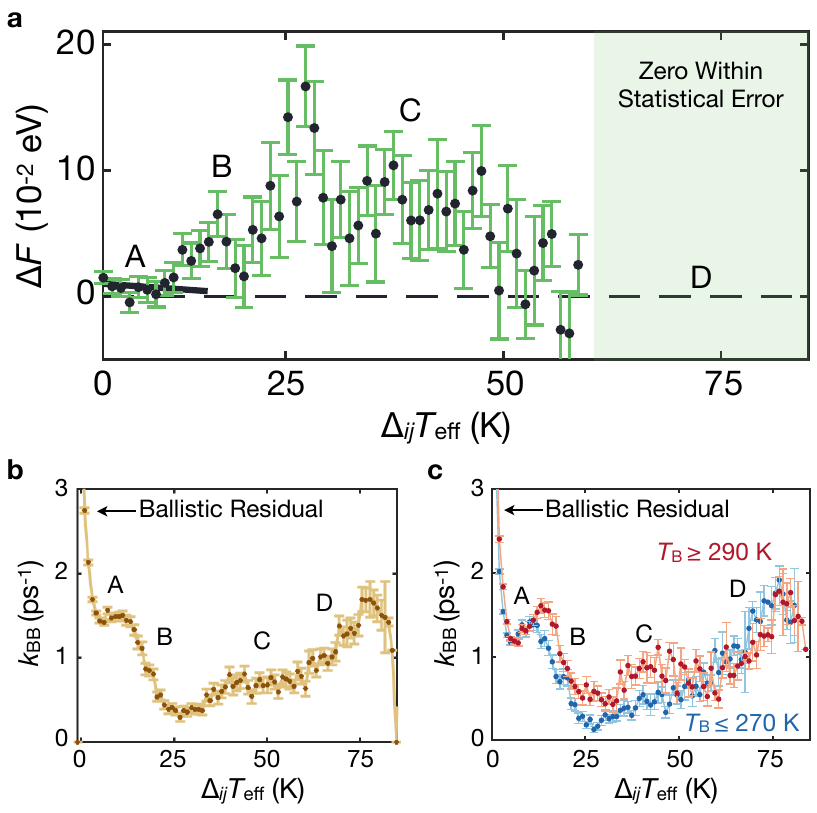} 
\caption{{\bf Transport barriers and kinetic parameters.} (a) Effective free energy barriers $\Delta F$ corresponding to different regions of the $J_\text{BB}$ flux profile.  Region A has nearly no barrier, but as the gradient becomes large, a barrier starts to form and increases in region B. In C, this barrier decreases until in D it is zero to within statistical errors (albeit, the uncertainty is large in this last region due to the limited number of samples for large temperature gradients, which occur only at short times). (b) Backbone rate distributions ($k_\text{BB}$)  for helical Aib$_{10}$ conformers.  Rate constants are parameterized by the temperature gradient $\Delta_{ij} T_\text{eff}$ between adjacent residues and (c) partitioned into low--temperature (blue; 230 K to 270 K) and high--temperature (red; 290 K to 330 K) regimes. The error bars are plus/minus one standard error.} \label{fig:rateprofile}
\end{figure}

{\bf Region B.} As $\Delta_{ij} T_\text{eff}$ goes above 15 K, the flux decreases with the increasing temperature gradient. This suggests the appearance of a vibrational mismatch between adjacent residues due to nonlinearity.  That is, adjacent residues separated by a sufficiently large temperature gradient will see different tiers of the energy landscape hierarchy and thus access different vibrational mode structures.  As a consequence, the molecular conformation is pushed into an activated region of the free energy landscape where the energy barrier is larger than the available kinetic energy and increases with $\Delta_{ij} T_\text{eff}$.  Moreover, the average temperature elevation does not substantially change for  $\Delta_{ij} T_\text{eff}$  in region B where the flux dips (Fig.~\ref{fig:fluxprofile}c).  Thus, barrier crossing is not aided by energy remaining from the initial deposition.  This is further supported by the separation of low-- and high--temperature curves, indicating that transport  increases with temperature --- a signature of a free energy barrier. The characteristic barriers can be estimated from the ratio of high-- and low--temperature fluxes (or rates), $J_\text{H}/J_\text{L} \approx 1.2 \approx \exp (-\Delta F/k_\text{B} T_\text{H}+\Delta F/k_\text{B} T_\text{L})$, giving values of $\Delta F$ that span from  $28$ meV to $167$ meV when we use the average temperature in each regime (i.e., $T_\text{L}=250$ K and $T_\text{H}=310$ K).   These effective barriers are precisely the energy scale leading to conformational changes that restore efficacious vibrational coupling. 

{\bf Region C.} As $\Delta_{ij} T_\text{eff}$ increases beyond 30 K, there is a substantial increase in flux for both low-- and high--temperature structures. In this case, a large $\Delta_{ij} T_\text{eff}$ implies a larger average temperature elevation for a given residue pair  (Fig.~\ref{fig:fluxprofile}c), as large gradients are primarily found at early times (and near the heater site) when a substantial fraction of initially deposited energy is present (Fig.~\ref{fig:fluxprofile}d). If we assume $\Omega$ remains the same, the temperature elevation $\Delta T_{\text{B},j}$ is enough to once again put transport in a stable regime of the landscape at this level of hierarchy, with a typical barrier energy of $67$ meV.  This yields an approximately linear region for $J_\text{BB}$ with a diffusivity $D_{\text{eff},\text{C}} = 1.9 \times 10^{-2}$~nm$^2$~ps$^{-1}$ ($\kappa_{\text{eff},\text{C}} = 3.2 \times 10^{-3}$~eV~K$^{-1}$~ps$^{-1}$).

{\bf Region D.} Increasing $\Delta_{ij} T_\text{eff}$ even further, beyond 50 K, leads to a  transport region with a larger diffusivity $D_{\text{eff},\text{D}} = 8.0 \times 10^{-2}$~nm$^2$~ps$^{-1}$ ($\kappa_{\text{eff},\text{D}} = 1.3 \times 10^{-2}$~eV~K$^{-1}$~ps$^{-1}$), corresponding to  over--the--barrier diffusion.  In this case, a new level of the energy landscape hierarchy becomes accessible, which would otherwise require strong activation at lower energies.

Figure~\ref{fig:rateprofile}a shows the effective free energy barriers in the different regimes, which are also reflected in the backbone rate constants (Fig.~\ref{fig:rateprofile}b,c). The $k_\text{BB}$ initially decrease with  $\Delta_{ij} T_\text{eff}$ (from 0 K to 4 K) due to a diminishing residual ballistic component when averaging at $\Delta t = 100$~fs. Overestimation of this signature (e.g., through an improper coarse--graining scale), can lead to the  discrepancies with experiment found in earlier theoretical analyses \cite{Botan2007,Nguyen2010}.  This is followed by a plateau in $k_\text{BB}$  at about 1.5 ps$^{-1}$ between 4 K to 15 K, followed by a drop as the landscape is pushed into a new, barrier--dominated region. After this, though, the larger $\Delta_{ij} T_\text{eff}$ correspond to a larger temperature elevation, bringing the events above the features in the energy landscape and raising $k_{BB}$ further.  \{Our methods extract the dependence on the local temperature gradients and, by spatiotemporal correlation, the temperature elevation. Beyond $\Delta_{ij} T_\text{eff} = 77$ K, the rate constants and fluxes decline sharply, reflecting very early dynamics where strong dynamical localization processes dominate.  These barriers collectively define the energy scales, and thus the rate of diffusion in conformational space \cite{Best2006}, that is associated with the mechanical dynamics of heat propagation at different temperatures.

\vspace{0.5\baselineskip}
\noindent{\bf Discussion}
\vspace{0.5\baselineskip}

While our NEMD simulations support that a transition~\cite{Botan2007,Backus2008b,Backus2009,Nguyen2010} in diffusivity is present, they do not support that the transition happens solely due to the existence of energy barriers, as stated in Ref. \cite{Botan2007,Backus2008,Kobus2010}, or glassy dynamics (which is certainly the case but does not pinpoint the particular processes that occur here). Rather, the transition is due to the development of region C physics: Energy flow, which largely happens from 0 to 10 ps, is in the presence of large $\Delta_{ij} T_\text{eff}$ (see initial time, high gradient line in Fig.~\ref{fig:fluxprofile}d) on top of equilibrium fluctuations ($\Delta_{ij} T_\text{eff} \approx \pm 10$ K).   We interpret this to indicate that large gradients give a vibrational mismatch via {\em nonlinear energy localization}, introducing a barrier to energy transport.  In this context, localization would then mediate the transition into a higher diffusivity regime --- thereby suggesting an origin of the sharpness of the transition. The increase of the base temperature reduces the vibrational mismatch by pushing the dynamics onto a different level of the landscape hierarchy. Simultaneous Arrhenius activation and barrier reduction conspire to give a sharp transition. More extensive simulations are necessary to make this precise. 

These findings demonstrate that energy transport gives quantitative information regarding the  biomolecular free energy landscape, its nonlinearity, and overall connectivity. Going beyond what we present here, the experimental analogues of our simulations offer potential probes of structural transitions, where a temperature--dependent change in the transport profile is a manifestation of the graph--theoretic topology associated with molecular contacts and nonlinear interactions of the dominant conformer(s).   In other words, thermal transport can be employed to devise `tomographies' that provide a complementary mapping of biomolecular structure, conformational dynamics, and folding pathways. While dominated by local contacts and secondary structure  within the simple Aib$_{10}$ peptide, we expect higher aspects of fold (tertiary, quaternary) to define these dynamics in increasingly complex biomolecules. Furthermore, such probes might excel for highly fluctuating systems such as intrinsically disordered proteins (IDPs), where efficacious thermal transport may still persist (addressed in the Supplementary Discussion), or as a means to dissect local shifts in vibrational mode structure during molecular signaling or allostery. These dynamics have been impervious to other spectroscopies.  Our approach provides the conceptual foundations and analysis tools that are directly applicable to experimental data, permitting the immediate interpretation of measurements that leverage local vibrational thermometry. In addition to the functional implications, the approach will also enable the development of a better understanding of what interactions look like at the atomic scale, and therefore better force-fields, and facilitate the design of nanodevices with directed, environmentally responsive heat transport mechanisms.

\vspace{0.1cm}
{\bf\noindent Methods}
\vspace{0.1cm}

\noindent {\bf Molecular dynamics simulations.} Our simulations consist of a modified Aib$_{10}$ peptide (AcOHN-Aib$_{10}$-COOCH$_3$), embedded in a box of 922 chloroform molecules.  Equilibration and ensemble generation are described in Ref. \cite{Elenewski2018}.  Prior to NEMD runs, structures are further equilibrated for 100~ps at each base ($T_\text{B}$) temperature (NPT; time step $\delta t = 1.0$~fs) followed by a 50~ps run with shorter time step (NPT; $\delta t = 0.1$~fs).  Using the final configurations, NEMD  (NVT; $\delta t = 0.1$~fs)  is initiated by heating the first residue of Aib$_{10}$ to $T_\text{B} + \Delta T$ ($\Delta T = 670$~K) for 1~ps, while holding the remaining atoms at $T_\text{B}$.  Thermostatting is then disabled and heat propagation monitored in the microcanonical ensemble.  Similar thermostatting protocols have been established as surrogates for explicit photoexcitation \cite{Nguyen2010,Buchenberg2016}.  NVT simulations employ a velocity Verlet integrator and modified Nos\'{e}--Hoover thermostat (damping = 100~fs), while NPT runs add a Martyna--Tobias--Klein barostat (damping = 1000~fs, eight member chain) \cite{Parrinello1981,Martyna1994,Shinoda2004}.   Isotropic cell fluctuations are allowed for NPT runs and initial velocities are assigned according to a Gaussian distribution.  Simulations employ CHARMM36 force field parameters \cite{MacKerell2004,Best2012}, CHARMM pair potentials (without CMAP parameters, as rationalized in Ref. \cite{Elenewski2018}), transferrable parameters for CHCl$_3$ \cite{Norbeg1998}, PPPM electrostatics (force cutoff $6.95 \times 10^{-3}$ pN; pair coupling rescaled at 1.0~nm, terminated at 1.35~nm) and the LAMMPS codebase \cite{Plimpton1995}.   We have adopted a thermostat  timescale that is faster than backbone amide relaxation and azobenzene isomerization in order to preserve transport--relevant dynamics.  While a slight  overpopulation of long--range modes remains possible, it would only serve to underestimate the impact of nonlinear localization while overestimating ballistic signatures  --- thus leaving our conclusions unaffected.

\vspace{\baselineskip}

\noindent {\bf Kinetic  fitting.}  While physically descriptive, the master equation, Equation (\ref{mastereq}), is underdetermined when fitting the simulated transport profiles $ E_j(t) = 3/2 N_j k_\text{B}  \langle T(t)\rangle $ for the $N_j$ atoms of the $j^\mathrm{th}$ residue.  As a simplifying approximation, we relate forward and reverse rate constants $k_{ij} = (f_i / f_j) k_{ji}$ through the degrees of freedom of each residue $f_j$, as required for detailed balance to hold at equilibrium.  We also restrict analysis to structurally homogeneous (helical) conformers, where the rate constants for hydrogen bond energy transfer  $k_{j,j+3} \approx k_{j,j+4} \approx k_\text{HB}$ and solvent coupling $k_{\text{s},j} \approx R_j \, k_\text{s}$  as can be approximated as uniform  (up to a fixed geometric factor $R_j$ for the surface area of terminal residues).  Under these conditions, we may fit the time dependence of the solvent $k_\text{s} \rightarrow k_\text{s}(t)$ and peptide rate constants,  $k_{ij} \rightarrow k_{ij}(t)$ and $k_\text{HB} \rightarrow k_\text{HB}(t)$, to account for the local temperature (which changes in time).  This is in contrast to prior efforts that assume a uniform and time--independent backbone rate constant $k_{j,j+1} = k_\text{BB}$  \cite{Buchenberg2016}.  

\par Rate constants $\mathbf{k}_j = (k_{1,2}, \dots, k_{N-1,N}, k_\text{H})$ at the $n^\mathrm{th}$ simulation time step are estimated for the linear system of Equation (\ref{mastereq}) though a constrained optimization
\begin{equation}
 \mathbf{k}(t_n) = \min_{\mathbf{k} \geq 0} \frac{1}{2} \vert\vert \mathbf{G}(t_n) \cdot \mathbf{k} - \mathbf{d}(t_n) \vert\vert^2
\end{equation}

\noindent where  $\mathbf{d}_j(t_n) = [E_j(t_n) - E_j(t_{n-1})] + k_\text{s}(t_n) [E_j(t_n) - E_\text{s}]$ captures energy redistribution among residues of the peptide.   The matrix $\mathbf{G}(t)$ is similarly defined so that $\mathbf{G}_{i,j}(t) = -\mathbf{G}_{i+1,j}(t) = -[E_i (t) - E_j(t) ]$ accommodates backbone energy transport and $\mathbf{G}_{i,N}(t)  = \sum_\ell [E_i - E_\ell]$ describes its hydrogen bonding counterpart to the $i^\mathrm{th}$ residue.  The solvent coupling rate  $k_\text{s}(t_n) = \sum_j [E_j(t_n) - E_j(t_{n-1})] / [E_j(t_n) - E_\text{s}(t_{n})]$ is then given by the energy exchanged between the peptide and the solvent at each time step (the solvent bath energy $E_\text{s}(t_{n}) = 3 N_j k_\text{B} T_\text{B} / 2$ is treated a constant).  

\vspace{0.1cm}
{\bf\noindent Data availability}
\vspace{0.1cm}

The authors declare that all data supporting the findings in this manuscript are available within the paper and its supplementary information.

\vspace{0.1cm}
{\bf\noindent Acknowledgements}
\vspace{0.1cm}

The authors would like to thank Thomas LeBrun for his insightful comments.  J. E. acknowledges support under the Cooperative Research Agreement between the University of Maryland and the National Institute for Standards and Technology Physical Measurement Laboratory, Award 70NANB14H209, through the University of Maryland.  K. V. was supported by the U.S. Department of Energy through the LANL/LDRD Program.  Computing resources were made available through the Los Alamos National Laboratory Institutional Computing Program, which is supported by the U.S. DOE National Nuclear Security Administration under contract no. DE-AC52-06NA25396, as well as the Maryland Advanced Research Computing Center (MARCC).

\vspace{0.1cm}
{\bf\noindent Author contributions}
\vspace{0.1cm}

J. E. performed the simulations and analysis. J. E. and M. Z. formulated the theoretical concepts. J. E., K. V., and M. Z. all contributed to the development of the ideas and preparation of the manuscript.

\vspace{0.1cm}
{\bf\noindent Competing interests}
\vspace{0.1cm}

The authors declare no competing interests.


\end{document}


\newcommand{\todo}[1] {{\color{black} #1}}
\newcommand{\LRC}{$L \leftrightarrow R$}
\newcommand{\I}{\mathcal{I}}
\newcommand{\E}{\mathcal{E}}
\newcommand{\C}{\mathcal{C}}
\newcommand{\OO}{\mathcal{O}}
\newcommand{\R}{\mathcal{R}}
\newcommand{\SC}{\mathcal{S}}

\begin{center}
{\bf Supplementary Information for \\ ``Topology, Landscapes, and Biomolecular Energy Transport''}\\
\vspace{\baselineskip}
Elenewski, \emph{et al.}\\

\end{center}

\vspace{5\baselineskip}
\clearpage

\section{Supplementary Figures}

\vspace{2\baselineskip}

\vspace{2\baselineskip}
\begin{figure}[h]
\bgroup
\setlength\tabcolsep{8.0pt}
\begin{tabular}{ccc}
\includegraphics[scale=0.95]{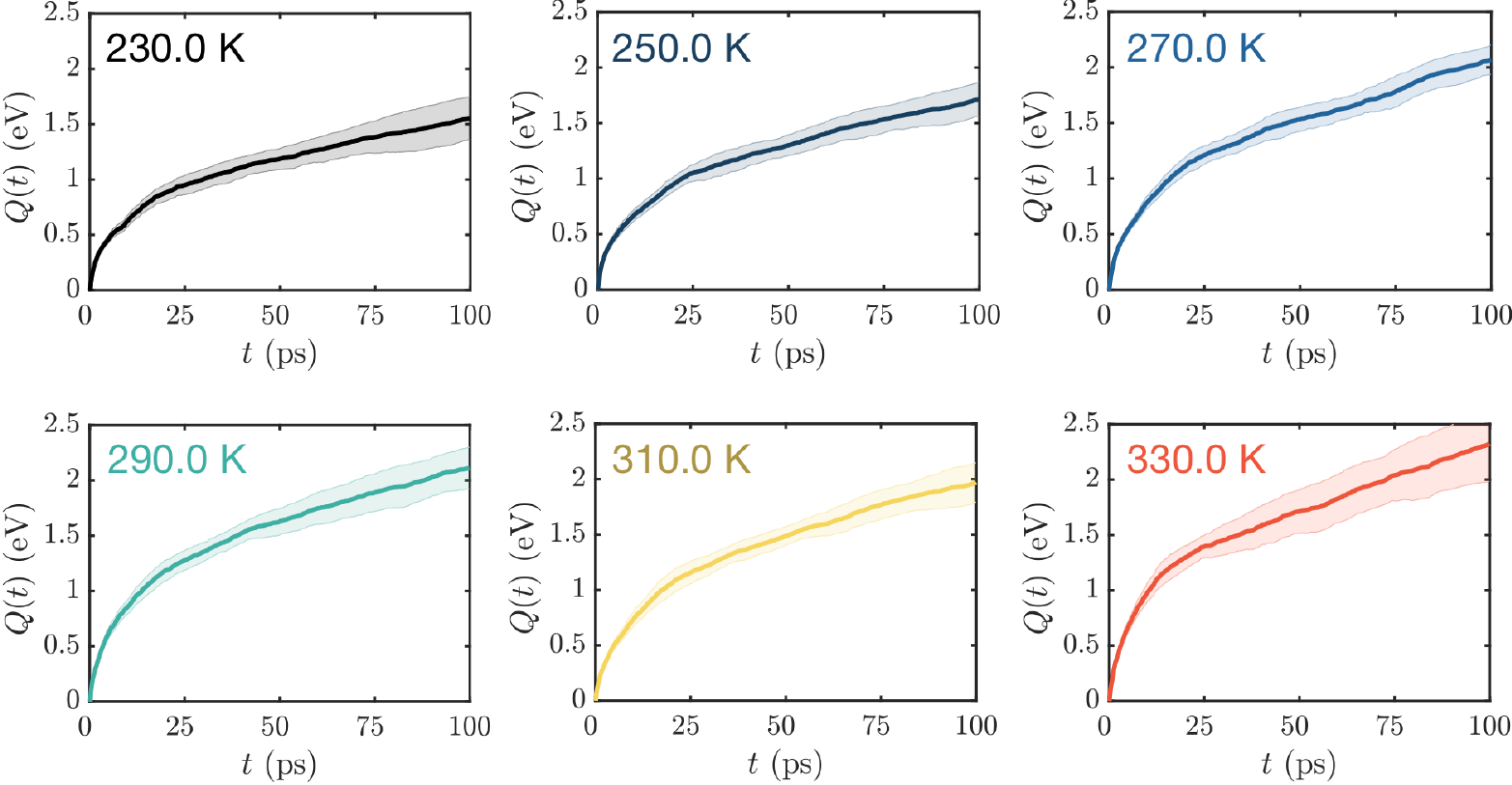}
\end{tabular}
\egroup
\justify { \bf \todo{Supplementary Figure 1.}  \bf Cumulative Heat Transfer Between Residues 1 and 2.} Energy transfer is quantified by integrating the heat flux \todo{$[J_\text{BB}]_{i,j}$} between the $i=1$ and $j=2$ residues up to time $t$ in the MD simulation trajectory: $Q(t) = \int_0^t J_{1,2} (t') \, dt'$.   Fluxes are obtained using the master equation analysis described in the parent manuscript, with time series data  coarse grained at $\Delta t$ = 100 fs prior to analysis.  Transfers exceed the 1.6 eV of heat added to the first residue due to residual error from short timescale transients.  This behavior may be mitigated by coarse graining the trajectory over larger temporal windows, at the cost of weaker statistics.  The bath temperature for each trace (\todo{$\Delta T_\text{B}$}) is indicated in the upper left hand side of the plot.   The error bands are plus/minus one standard error.  
\label{fig:fitComponent}
\end{figure}

\clearpage
\vspace{2\baselineskip}
\begin{figure}[h]
\bgroup
\setlength\tabcolsep{8.0pt}
\begin{tabular}{ccc}
\includegraphics[scale=0.95]{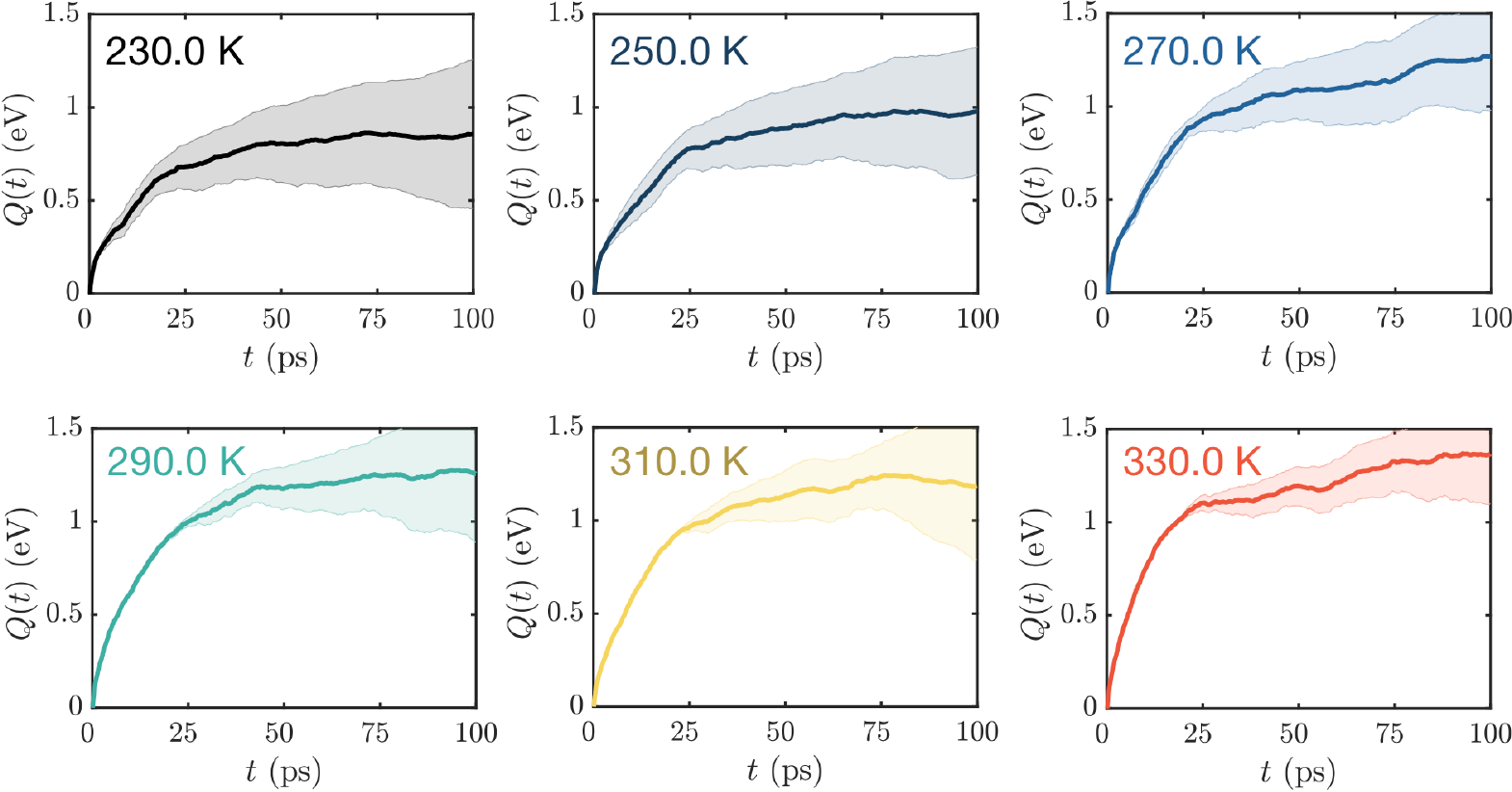}
\end{tabular}
\egroup
\justify {\bf\todo{Supplementary Figure 2.}} {\bf Cumulative Heat Transfer Between Residues 2 and 3.} Energy transfer is quantified by integrating the heat flux \todo{$[J_\text{BB}]_{i,j}$} between the  $i=2$ and $j=3$ residues up to time $t$ in the MD simulation trajectory: $Q(t) = \int_0^t J_{2,3} (t') \, dt'$.   Fluxes are obtained using the master equation analysis described in the parent manuscript, with time series data  coarse grained at $\Delta t$ = 100 fs prior to analysis.  The bath temperature for each trace (\todo{$\Delta T_\text{B}$}) is indicated in the upper left hand side of the plot.  The error bands are plus/minus one standard error.  
\label{fig:fitComponent}
\end{figure}

\clearpage
\vspace{2\baselineskip}
\begin{figure}[h]
\bgroup
\setlength\tabcolsep{8.0pt}
\begin{tabular}{ccc}
\includegraphics[scale=0.95]{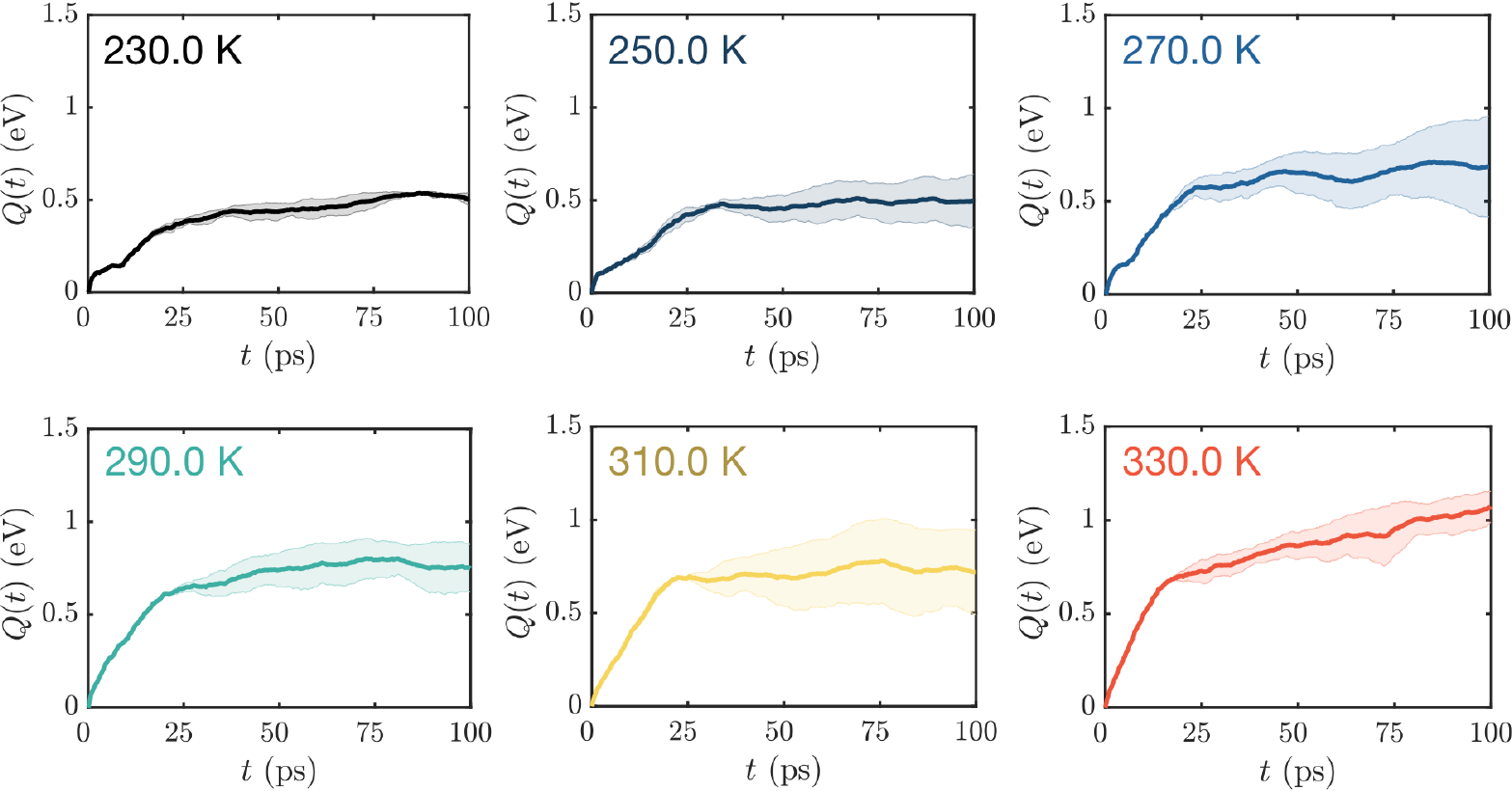}
\end{tabular}
\egroup
\justify {\bf\todo{Supplementary Figure 3.}} {\bf Cumulative Heat Transfer Between Residues 3 and 4.} Energy transfer is quantified by integrating the heat flux \todo{$[J_\text{BB}]_{i,j}$} between the  $i=3$ and $j=4$  residues up to time $t$ in the MD simulation trajectory: $Q(t) = \int_0^t J_{3,4} (t') \, dt'$.   Fluxes are obtained using the master equation analysis described in the parent manuscript, with time series data  coarse grained at $\Delta t$ = 100 fs prior to analysis. The bath temperature for each trace (\todo{$\Delta T_\text{B}$}) is indicated in the upper left hand side of the plot.   The error bands are plus/minus one standard error. 
\label{fig:fitComponent}
\end{figure}

\clearpage
\vspace{2\baselineskip}
\begin{figure}[h]
\bgroup
\setlength\tabcolsep{8.0pt}
\begin{tabular}{ccc}
\includegraphics[scale=0.95]{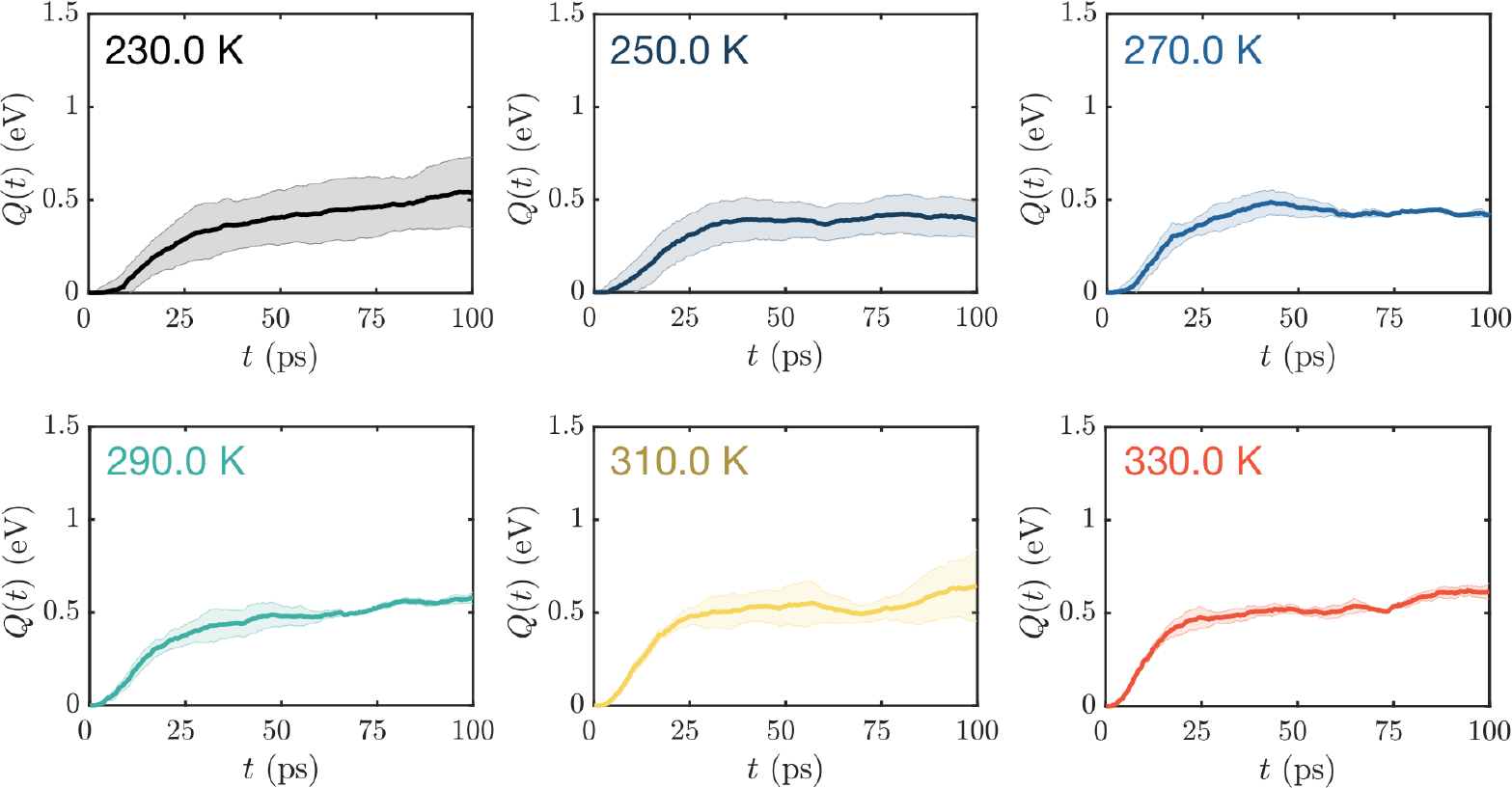}
\end{tabular}
\egroup
\justify {\bf\todo{Supplementary Figure 4.}} {\bf Cumulative Heat Transfer Between Residues 4 and 5.} Energy transfer is quantified by integrating the heat flux \todo{$[J_\text{BB}]_{i,j}$} between the  $i=4$ and $j=5$  residues up to time $t$ in the MD simulation trajectory: $Q(t) = \int_0^t J_{4,5} (t') \, dt'$.   Fluxes are obtained using the master equation analysis described in the parent manuscript, with time series data  coarse grained at $\Delta t$ = 100 fs prior to analysis. The bath temperature for each trace (\todo{$\Delta T_\text{B}$}) is indicated in the upper left hand side of the plot. The error bands are plus/minus one standard error.  
\label{fig:fitComponent}
\end{figure}

\clearpage
\vspace{2\baselineskip}
\begin{figure}[h]
\bgroup
\setlength\tabcolsep{8.0pt}
\begin{tabular}{ccc}
\includegraphics[scale=0.95]{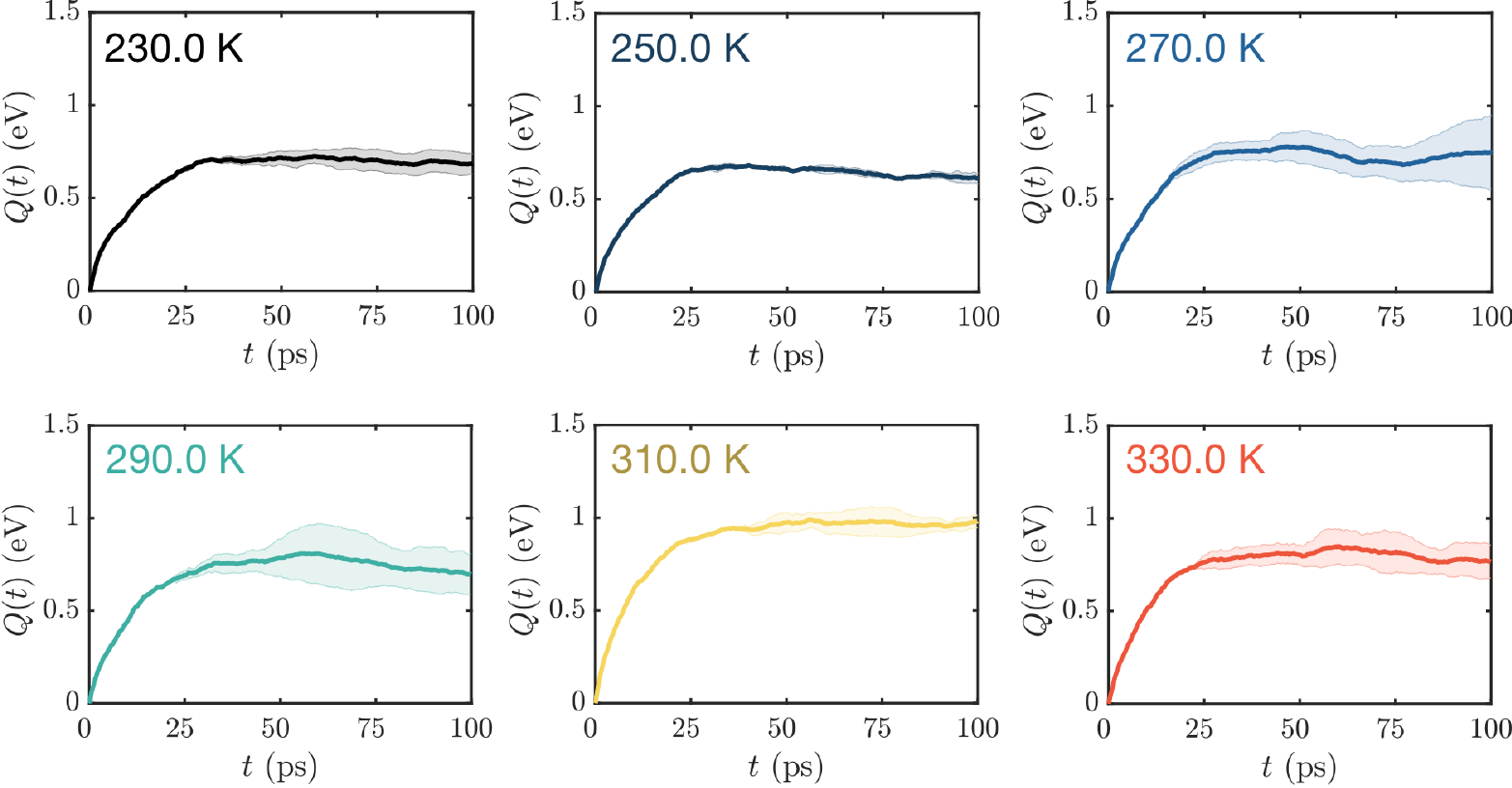}
\end{tabular}
\egroup
\justify {\bf\todo{Supplementary Figure 5.}} {\bf Cumulative Heat Transfer Between Residues 5 and 6.} Energy transfer is quantified by integrating the heat flux \todo{$[J_\text{BB}]_{i,j}$} between the  $i=5$ and $j=6$ residues up to time $t$ in the MD simulation trajectory: $Q(t) = \int_0^t J_{5,6} (t') \, dt'$.   Fluxes are obtained using the master equation analysis described in the parent manuscript, with time series data  coarse grained at $\Delta t$ = 100 fs prior to analysis. The bath temperature for each trace (\todo{$\Delta T_\text{B}$}) is indicated in the upper left hand side of the plot. The error bands are plus/minus one standard error.   
\label{fig:fitComponent}
\end{figure}

\clearpage
\vspace{2\baselineskip}
\begin{figure}[h]
\bgroup
\setlength\tabcolsep{8.0pt}
\begin{tabular}{ccc}
\includegraphics[scale=0.95]{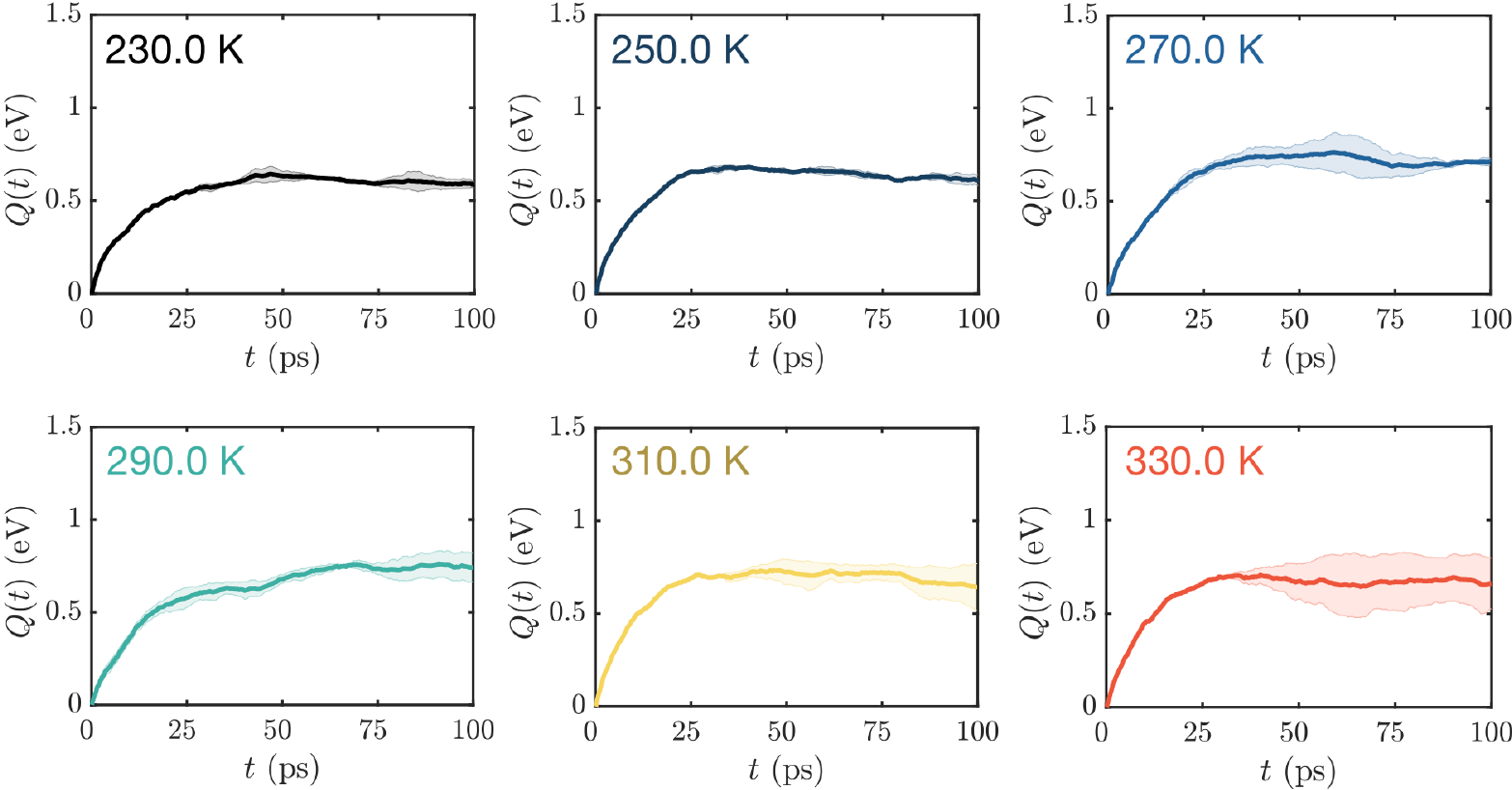}
\end{tabular}
\egroup
\justify {\bf\todo{Supplementary Figure 6.}} {\bf Cumulative Heat Transfer Between Residues 6 and 7.} Energy transfer is quantified by integrating the heat flux \todo{$[J_\text{BB}]_{i,j}$} between the $i=6$ and $j=7$ residues up to time $t$ in the MD simulation trajectory: $Q(t) = \int_0^t J_{6,7} (t') \, dt'$.   Fluxes are obtained using the master equation analysis described in the parent manuscript, with time series data  coarse grained at $\Delta t$ = 100 fs prior to analysis. The bath temperature for each trace (\todo{$\Delta T_\text{B}$}) is indicated in the upper left hand side of the plot.  The error bands are plus/minus one standard error.  
\label{fig:fitComponent}
\end{figure}

\clearpage
\vspace{2\baselineskip}
\begin{figure}[h]
\bgroup
\setlength\tabcolsep{8.0pt}
\begin{tabular}{ccc}
\includegraphics[scale=0.95]{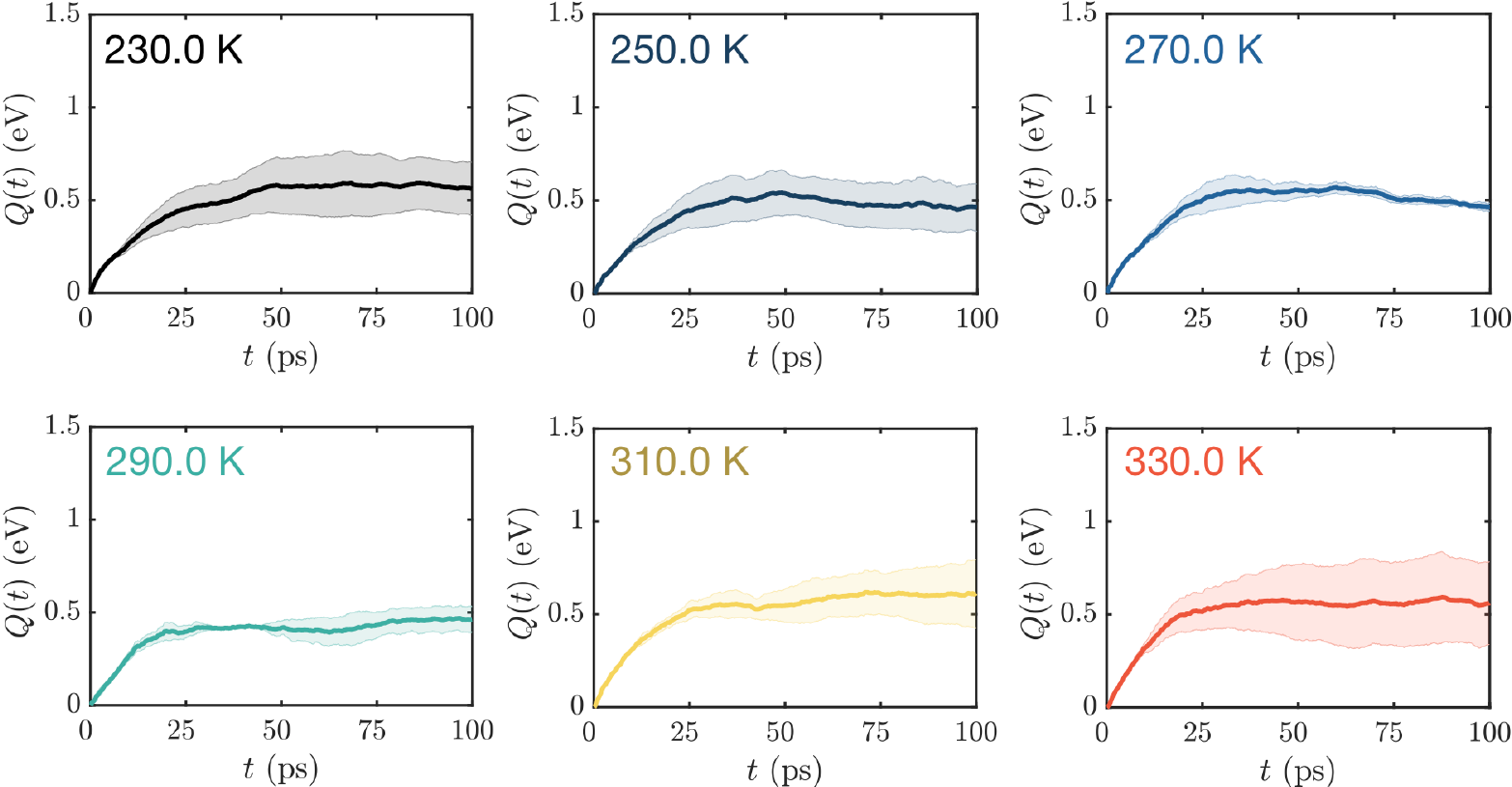}
\end{tabular}
\egroup
\justify {\bf\todo{Supplementary Figure 7.}} {\bf Cumulative Heat Transfer Between Residues 7 and 8.} Energy transfer is quantified by integrating the heat flux \todo{$[J_\text{BB}]_{i,j}$} between the  $i=7$ and $j=8$ residues up to time $t$ in the MD simulation trajectory: $Q(t) = \int_0^t J_{7,8} (t') \, dt'$.   Fluxes are obtained using the master equation analysis described in the parent manuscript, with time series data  coarse grained at $\Delta t$ = 100 fs prior to analysis. The bath temperature for each trace (\todo{$\Delta T_\text{B}$}) is indicated in the upper left hand side of the plot. The error bands are plus/minus one standard error.   
\label{fig:fitComponent}
\end{figure}

\clearpage
\vspace{2\baselineskip}
\begin{figure}[h]
\bgroup
\setlength\tabcolsep{8.0pt}
\begin{tabular}{ccc}
\includegraphics[scale=0.95]{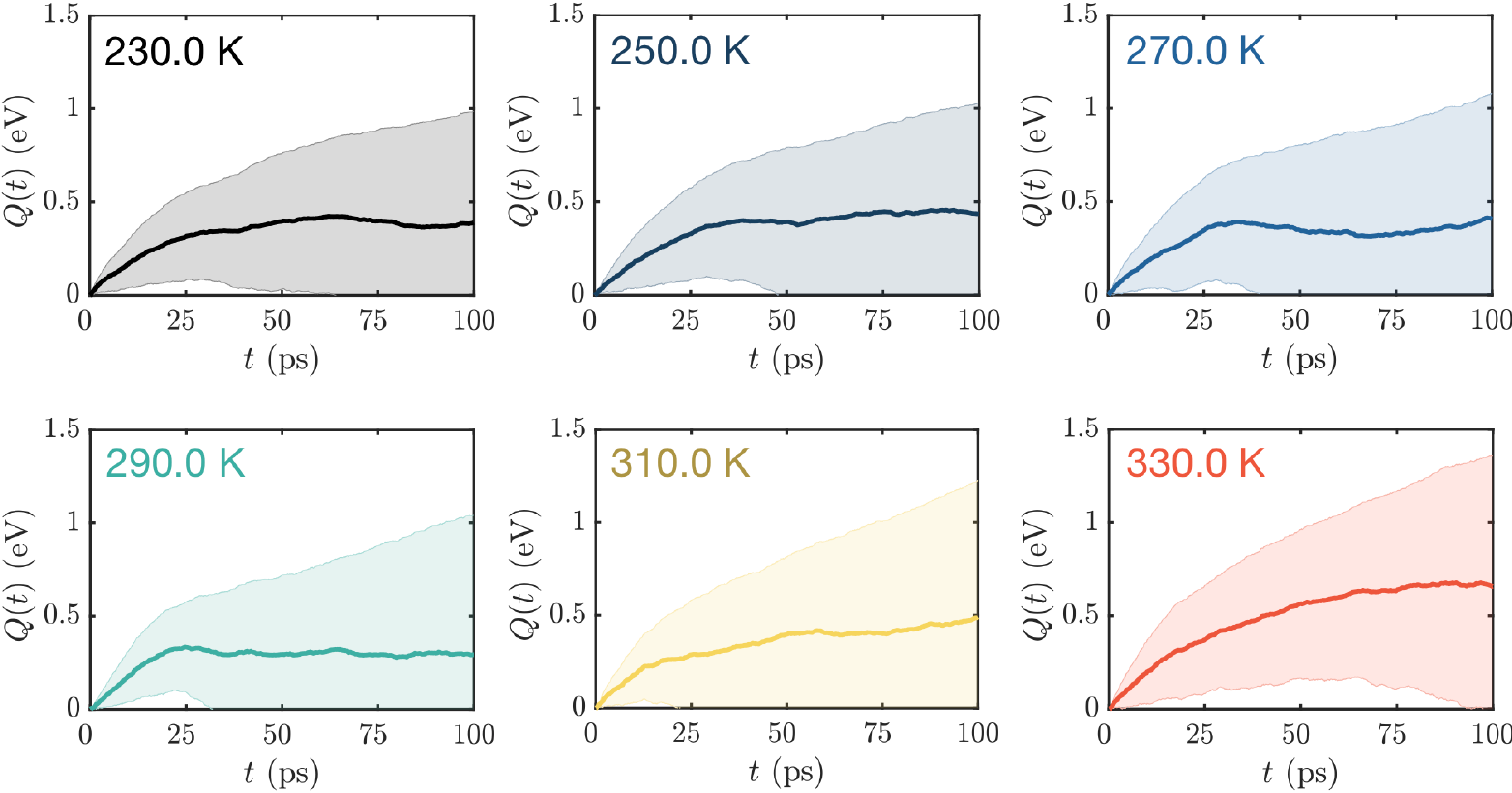}
\end{tabular}
\egroup
\justify {\bf\todo{Supplementary Figure 8.}}  {\bf Cumulative Heat Transfer Between Residues 8 and 9.} Energy transfer is quantified by integrating the heat flux \todo{$[J_\text{BB}]_{i,j}$} between the  $i=8$ and $j=9$ residues up to time $t$ in the MD simulation trajectory: $Q(t) = \int_0^t J_{8,9} (t') \, dt'$.   Fluxes are obtained using the master equation analysis described in the parent manuscript, with time series data  coarse grained at $\Delta t$ = 100 fs prior to analysis. The bath temperature for each trace (\todo{$\Delta T_\text{B}$}) is indicated in the upper left hand side of the plot. The error bands are plus/minus one standard error.  
\label{fig:fitComponent}
\end{figure}

\clearpage
\vspace{2\baselineskip}
\begin{figure}[h]
\bgroup
\setlength\tabcolsep{8.0pt}
\begin{tabular}{ccc}
\includegraphics[scale=0.95]{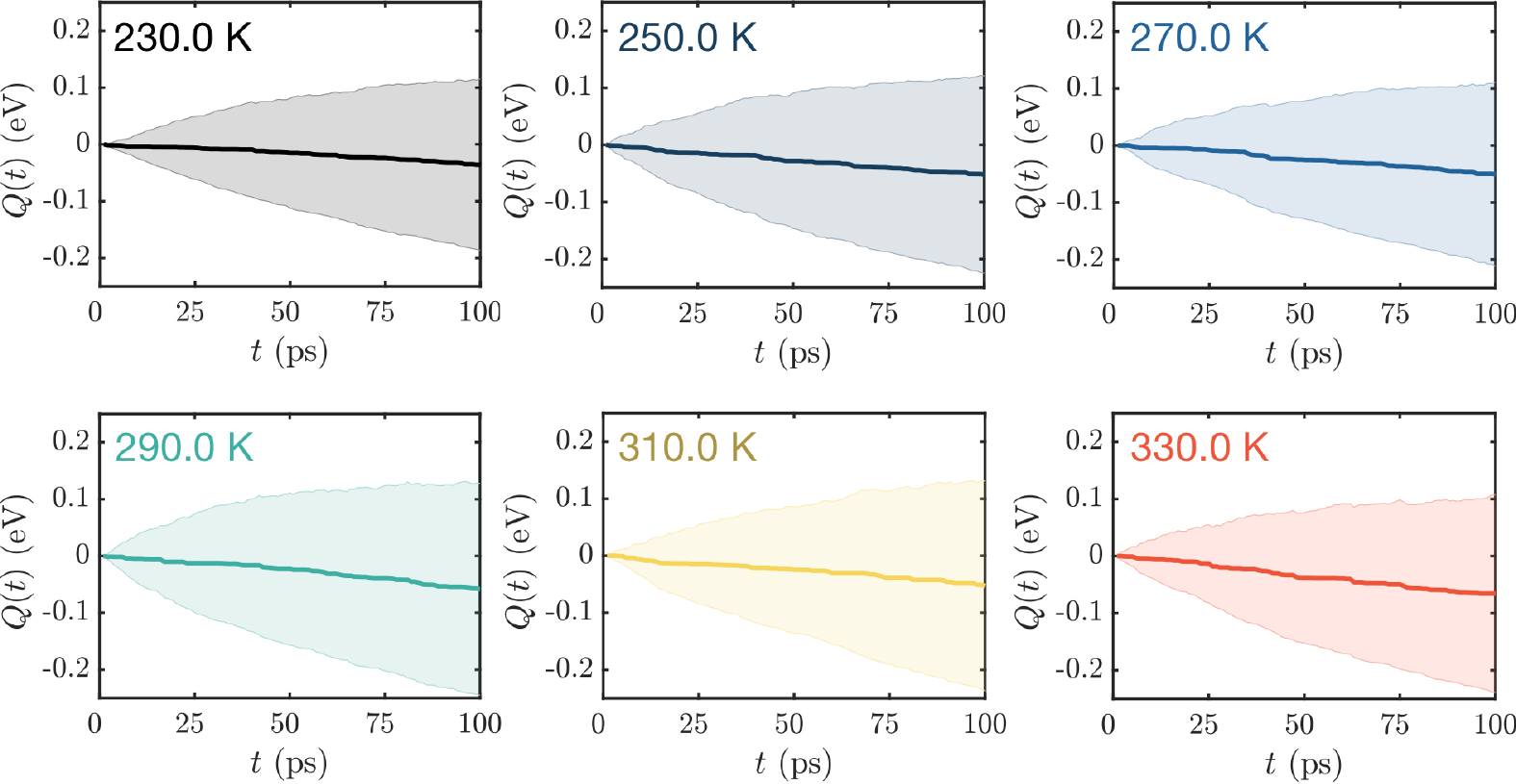}
\end{tabular}
\egroup
\justify {\bf\todo{Supplementary Figure 9.}}  {\bf Cumulative Heat Transfer Between Residues 9 and 10.} Energy transfer is quantified by integrating the heat flux \todo{$[J_\text{BB}]_{i,j}$} between the  $i=9$ and $j=10$ residues up to time $t$ in the MD simulation trajectory: $Q(t) = \int_0^t J_{9,10} (t') \, dt'$.   Fluxes are obtained using the master equation analysis described in the parent manuscript, with time series data  coarse grained at $\Delta t$ = 100 fs prior to analysis.  There is a small negative $Q(t)$. This is within the error, but may be due to the high flexibility of residue 10 leading to intermittent contact with the remainder of the peptide even with in the helical structure. This could give a thermal transport pathway from residue 10 to 9, or some residual effect from the solvent. The error bands are plus/minus one standard error.   
\label{fig:fitComponent}
\end{figure}

\clearpage

\begin{figure}[h]
\bgroup
\setlength\tabcolsep{8.0pt}
\includegraphics[scale=0.95]{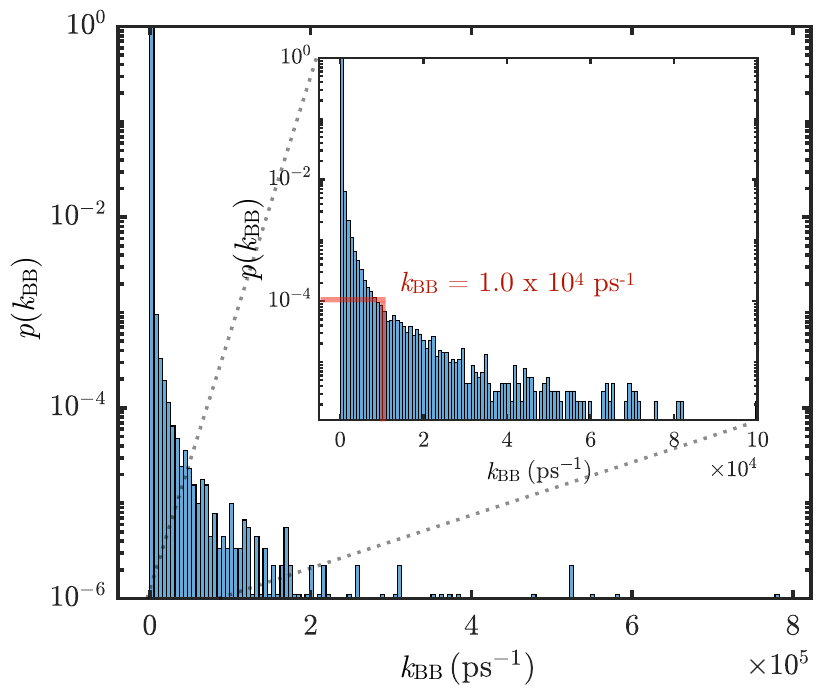}\\
\egroup
\justify { \bf \todo{Supplementary Figure 10.} Rate constant distribution.} Probability density distribution $\todo{p(k_\text{BB})}$ for backbone rate constants $\todo{k_\text{BB}}$, as obtained using the master equation--based fitting method (\todo{Equation 1} of the parent manuscript).  Physically extremal values, defined as rates that exceed $6.2 \times 10^{3}$ ps$^{-1}$, have a statistical weight of less than $1.0 \times 10^{-4}$, attesting to the robustness of the fitting algorithm.  The distribution is presented as an aggregate for helical conformers at all temperatures, with simulation time series data coarse grained over $\Delta t = 100$ fs intervals prior to analysis.  
\label{fig:fitComponent}
\end{figure}
\clearpage

\begin{figure}[h]
\bgroup
\setlength\tabcolsep{8.0pt}
\begin{tabular}{ccc}
\includegraphics[scale=0.95]{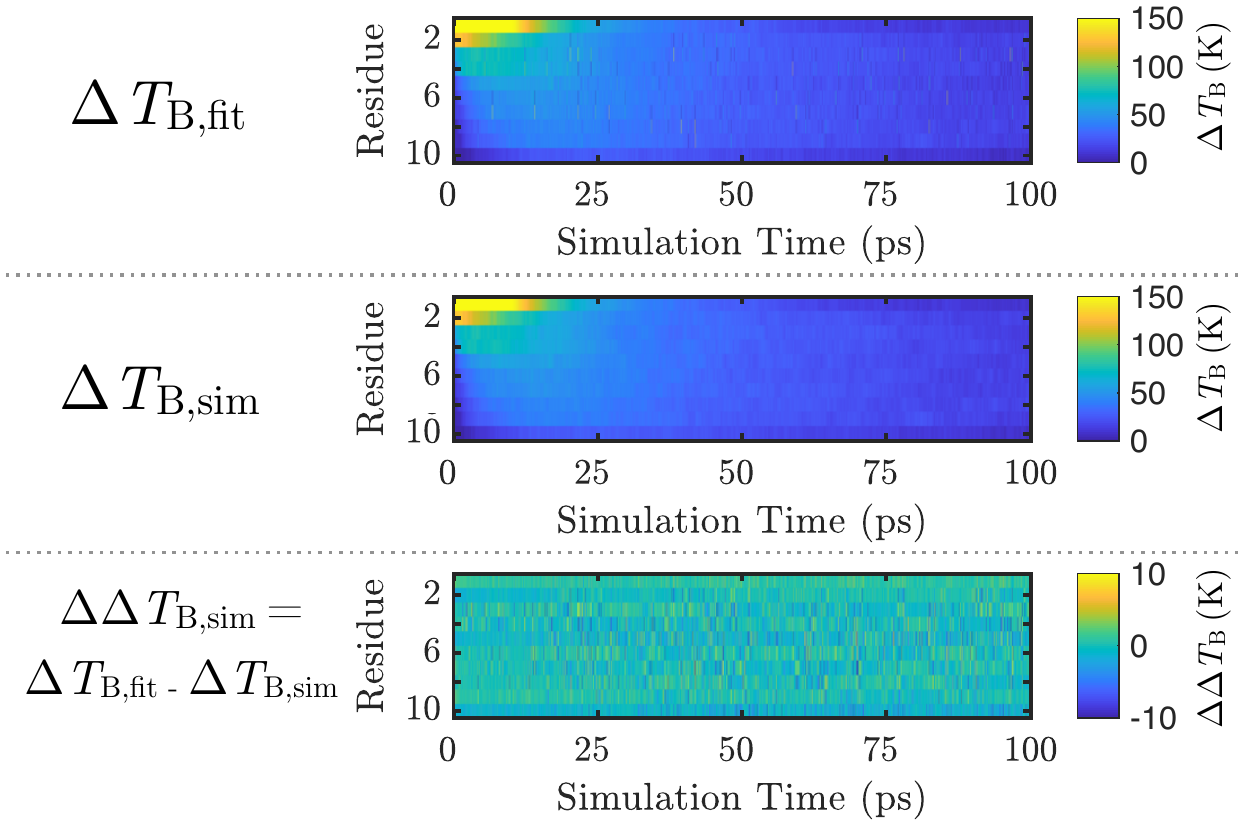}
\end{tabular}
\egroup
\justify { \bf \todo{Supplementary Figure 11.} Reconstructed thermal transport profiles.} To demonstrate robustness of our master equation reconstruction (\todo{Equation 1} of the manuscript), molecular dynamics simulation profiles $\todo{\Delta T_{\text{B},\text{sim}}}$ are propagated forward by one analysis timestep ($\Delta t = 100$ fs) using the backbone rate constants $\todo{k_\text{BB}}$ fit at that step.   In doing so, we generate a temperature elevation profile $\todo{\Delta T_{\text{B},\text{fit}}}$ for fit data.  The deviation between fit and simulation data $\todo{\Delta \Delta T_\text{B}} = \todo{\Delta T_{\text{B},\text{fit}}} - \todo{\Delta T_{\text{B},\text{sim}}}$ affords a metric for quality of reconstruction, exhibiting variations that are generally below $\pm 10.0$ K (\todo{Supplementary Figure 12}).  Data are presented for simulations at $\todo{T_\text{B}} = 230.0$ K, with MD simulation data coarse grained over $\Delta t = 100$ fs intervals prior to analysis.
\label{fig:fitComponent}
\end{figure}
\clearpage

\begin{figure}[h]
\bgroup
\setlength\tabcolsep{8.0pt}
\begin{tabular}{ccc}
\includegraphics[scale=0.95]{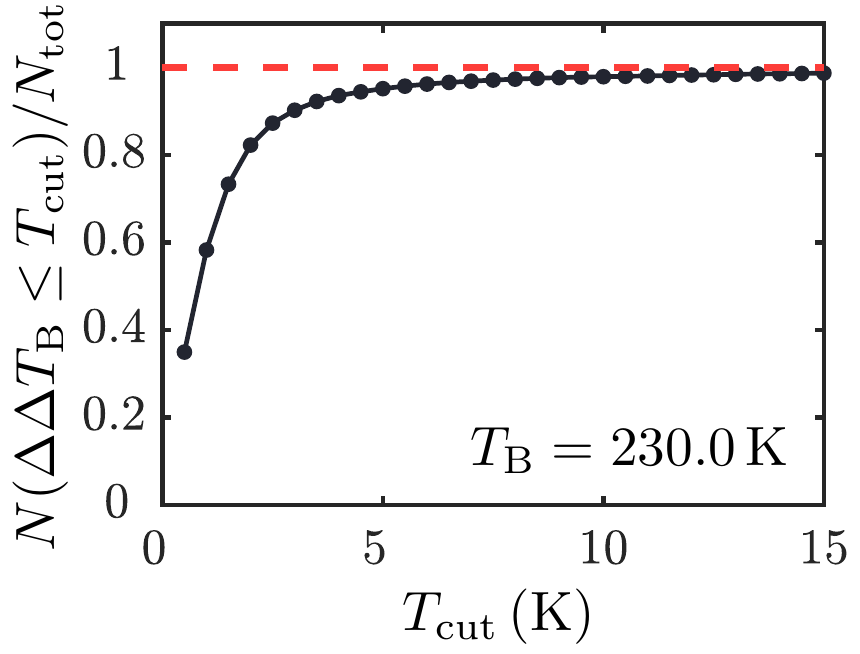}
\end{tabular}
\egroup
\justify { \bf \todo{Supplementary Figure 12.}  \bf Percentage of trajectory frame deviations by temperature cutoff.} The variation between fit rate profiles and raw MD simulation (\todo{Supplementary Figure 11}) may be further quantified through the percentage of trajectory propagations $N / N_\text{tot}$ with a temperature deviation $\todo{\Delta \Delta T_\text{B}} = \todo{\Delta T_{\text{B},\text{fit}}} -  \todo{\Delta T_{\text{B},\text{sim}}}$ lying at or below a given cutoff $\todo{\Delta \Delta T_\text{B}} \leq T_\text{cut}$ (here $N_\text{tot}$ is the total number of propagations).  In this case, 95.2 \% of propagations exhibit a deviation of less than 5.0 K over 100 fs, while 98.0 \% show a deviation of less than 10.0 K over the same time interval.  At deviations up to $T_\text{cut} = $ 15.0 K, we find that 99.0 \% of propagations will lie below the cutoff.  While not employed here, these values can be used to filter erroneous fits during analysis.  In this case, $T_\text{cut} = 10.0$ K would account for pathologies due to numerical instability while retaining robust counting statistics.  Data are presented for simulations at $\todo{T_\text{B}} = 230.0$, with MD simulation data coarse grained over $\Delta t = 100$ fs intervals prior to analysis.
\label{fig:fitComponent}
\end{figure}

\clearpage 

\vspace{2\baselineskip}

\begin{figure}[h]
\bgroup
\setlength\tabcolsep{8.0pt}
\begin{tabular}{ccc}
\includegraphics[scale=1.5]{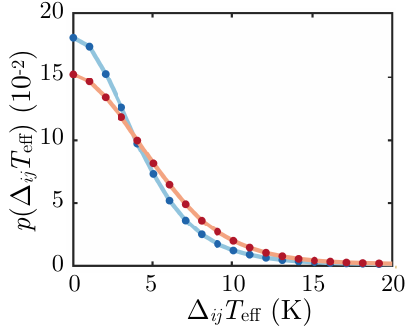}
\end{tabular}
\egroup
\justify {\bf\todo{Supplementary Figure 13.}}  {\bf Distribution of backbone temperature gradients.} The probability density distribution  $p(\Delta_{ij} T_\text{eff})$ for the effective temperature gradient $\Delta_{ij} T_\text{eff}$ between adjacent residues (taken as positive when $i > j$) is  calculated directly from simulation data.  Data are partitioned into  low--temperature (blue; 230 K to 270 K) and high--temperature (red; 290 K to 330 K) regimes.  Distributions are calculated using trajectory frames that have been coarse grained over $\Delta t = 100$ fs intervals, and are themselves averaged into 1.0 K bins.    
\label{fig:rateDistrib}
\end{figure}
\clearpage

\vspace{2\baselineskip}

\begin{figure}[h]
\bgroup
\setlength\tabcolsep{8.0pt}
\begin{tabular}{ccc}
\includegraphics[scale=1.5]{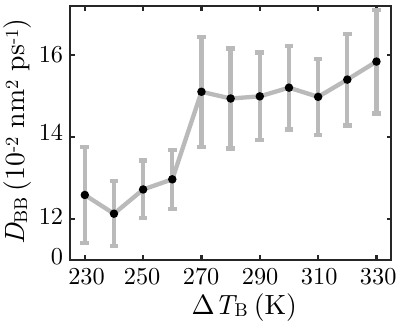}
\end{tabular}
\egroup
\justify {\bf\todo{Supplementary Figure 14.}}  {\bf Heat diffusivities calculated from backbone rate constants.}  The effective backbone heat diffusivity $\todo{D_\text{BB}} = \todo{k_\text{BB}}\, (\Delta x)^2 $ may be estimated using the  rate constants $\todo{k_\text{BB}}$ from the master equation analysis.  Diffusivities are calculated as $\todo{D_\text{BB}(T_\text{B})} = \sum_{\ell=1}^N \todo{k_{\text{BB},\ell}} \, p_\ell\,(\overline{\Delta x})^2$ at each bath temperature \todo{$T_\text{B}$}, where the summation is over all $N$ bins of the thermal gradient $\Delta_{ij} T_\text{eff}$ distribution, $p_\ell$ is the weight assigned to each bin (\todo{Supplementary Figure 13}), and $\overline{\Delta x}$ is the mean residue separation. The resulting diffusivities exhibit scaling that mimics the full simulation data (Fig.~2a in the manuscript), albeit with a slightly larger magnitude.  While  deviations exist (particularly for \todo{$T_\text{B}$} = 240 K), these likely reflect the simplified interactions accommodated by our model, alongside limitations due to sampling.    The overall similarity suggests that most critical processes are captured by  the master equation approach, supporting the scope of our interpretation.   The error bars are plus/minus one standard error.      
\label{fig:fitComponent}
\end{figure}
\clearpage

\begin{figure}[h]
\bgroup
\setlength\tabcolsep{8.0pt}
\begin{tabular}{ccc}
\includegraphics[scale=1.5]{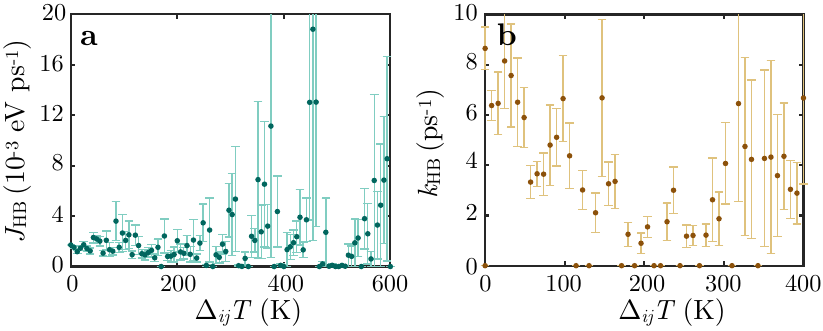}
\end{tabular}
\egroup
\justify {\bf\todo{Supplementary Figure 15.}}  {\bf Hydrogen Bond Rates and Fluxes.} (a) Hydrogen bond flux distributions (\todo{$J_\text{HB}$}) for helical Aib$_{10}$ conformers.   Fluxes are parameterized by the effective temperature gradient $\Delta_{ij} T_\text{eff}$ between adjacent residues, taken as positive when $i > j$. (b) Rate constants ($\todo{k_\text{HB}}$) corresponding to the hydrogen bond flux distribution.  Rates and fluxes are obtained by fitting a kinetic master equation (\todo{Equation~1} from the parent manuscript) to MD simulation data that has been coarse grained over $\Delta t = 100$~fs intervals.  The resulting fluxes and rates are block averaged into 1.0 K bins.   The error bars are plus/minus one standard error.     
\label{fig:fitComponent}
\end{figure}
\clearpage

\vspace{2\baselineskip}

\todo{\section{Supplementary Discussion}}

\par As a general rule, different folds of the Aib$_{10}$ peptide (or other biomolecules) have a conformationally--dependent solvent--accessible surface area (SASA).  Since thermal conduction between the peptide and the solvent bath occurs at this interface, it is expected that the overall thermalization profile will also depend on molecular conformation.   To quantify this, we exploit the largely diffusive nature of heat conduction in Aib$_{10}$ to construct a simple model for thermal relaxation.  In this case, we assume two compartments --- consisting of the peptide and the solvent --- which undergo strictly conductive heat transfer according to Fourier's law (i.e., no convective contribution).  Assuming that the solvent bath is much larger than the peptide, with minimal local solvent heating, we can write the  total kinetic energy content $E(t)$ of Aib$_{10}$ in the time--dependent form 

\begin{equation} 
E(t) = E_\text{B} + [E(0) - E_\text{B}] e^{-\todo{k_\text{c}} t},
\end{equation}

\noindent where $E_\text{B}$ is the net kinetic energy content of Aib$_{10}$ when in thermal equilibrium with the bath, $E(0)$ is the net kinetic energy content of peptide immediately following heating, and $\todo{k_\text{c}}$ is a characteristic time constant for heat transfer to the solvent.   Using this expression, we employ three NEMD ensembles, containing 750 simulations each (at $\todo{T_\text{B}} = 230.0$ K), for helix, hairpin, and completely extended Aib$_{10}$ conformers.  The fits resulting from  this protocol are depicted in \todo{Supplementary Figure 16} and summarized in \todo{Supplementary Table 1}.

\par Our simple cooling model is generally robust when applied to thermal relaxation in Aib$_{10}$.  Modest deviations between the resulting fits and simulation data are observed at early times ($t \leq 2.5$ ps) and high temperatures, where ballistic processes likely shunt heat to the solvent more rapidly than allowed by a diffusive mechanism.  Outside of this region, the thermal transport dynamics are relatively similar for the helix, hairpin, and extended coil, with the most prominent (relative) conformational variation observed for the cooling rate constant $\todo{k_\text{c}}$.  In our simulations, the extended coil dissipates heat most rapidly to solvent, followed by the structured $\alpha$--helix fold and comparatively globular hairpin conformations.  While statistically significant, this effect is also small --- largely due to the fact that Aib$_{10}$ dynamics are surface--dominated in any conformer due to finite--size effects.  

\par The overall scaling trend for $\todo{k_\text{c}}$ is mirrored when comparing the mean SASA between conformers (\todo{Supplementary Table 1}).  Quantitative differences nonetheless exist between SASA and $\todo{k_\text{c}}$ ratios [$\todo{k_\text{c}}(\text{coil}) / \todo{k_\text{c}}(\text{hairpin}) = 1.10$, while we find $\text{SASA}(\text{coil})/\text{SASA}(\text{hairpin}) = 1.26$], with the surface area contribution underestimated in the rate constants.  While a variety of factors may collude in this effect, the conformer--dependent heat transport rates within the peptide likely make the largest contribution.  That is, the helical and hairpin conformations possess high thermal diffusivities and auxiliary conduction pathways (due to molecular topology), which facilitate the transfer of heat to away from the heater and throughout the peptide (\todo{Supplementary Figure 17}).  Since the local cooling rate $\dot{E}_j (t)$ at the $j$--th residue is proportional to the local temperature gradient between the peptide and the solvent $\dot{E}_j (t) \propto T_j (t) - \todo{T_{\text{s},j}(t)}$,  this accelerates solvent relaxation.  The peptide thus acts as a radiator, relinquishing heat to the bath while circumventing local solvent heating --- and thus partially mitigating the insulating effect of molecular conformation.  We expect the contribution of this effect to be less in larger biomolecules --- in a manner that depends on the surface--to--volume ratio --- where redistribution pathways may also lead deeper into the molecular `bulk' and thus away from the solvent interface.

\clearpage 

\vspace{2\baselineskip}
\begin{figure}[h]
\bgroup
\setlength\tabcolsep{8.0pt}
\begin{tabular}{ccc}
\includegraphics[scale=0.95]{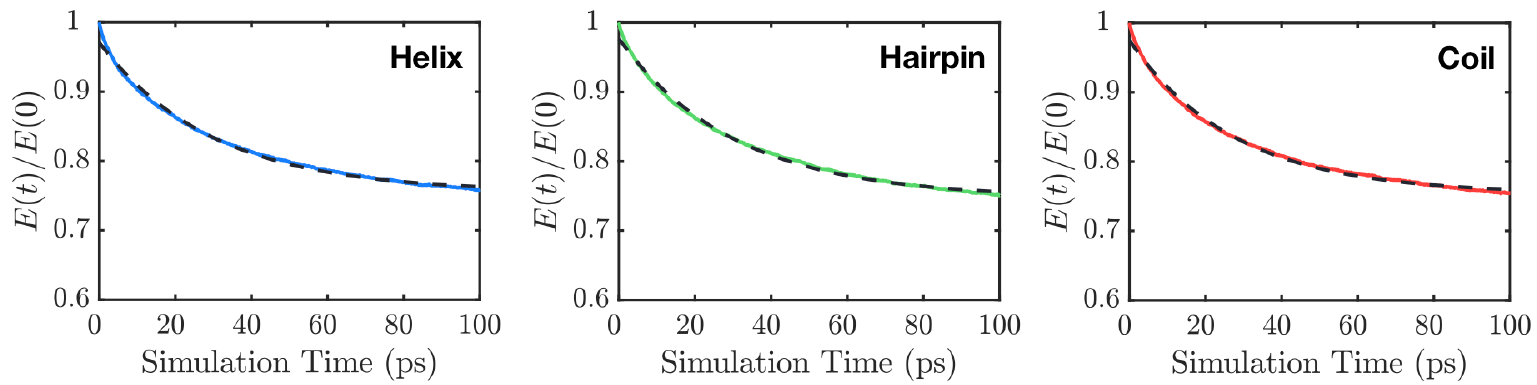}
\end{tabular}
\egroup
\justify {\bf\todo{Supplementary Figure 16.}} {\bf Conformation--dependent cooling.} Application of the the cooling model $E(t) = E_\text{B} + [E(0) - E_\text{B}] e^{-\todo{k_\text{c}} t}$ to NEMD simulations of distinct Aib$_{10}$ conformers at $\todo{T_\text{B}} = 230.0$ K.  Nonlinear least--squares fitting was performed to the mean kinetic energy profile of each NEMD ensemble (each containing 750 distinct conformations) with block averaging into 10 fs bins  prior to analysis.  Error bands corresponding to the block standard error are displayed alongside simulation data (colored lines; bands are on the order of the line width), while the fits are depicted as a black dashed line.  Data are measured relative to the peak net kinetic energy of Aib$_{10}$, with simulation methods identical to the parent manuscript.  
\label{fig:coolingRate}
\end{figure}

\clearpage

\begin{table}
\begin{center}
\begin{tabular}{ |l||c|c|c|c|}
\hline
 Conformer & $E_\text{B} / E(0)$    & $[E(0) - E_\text{B}] / E(0)$ & $\todo{k_\text{c}}$ (ps$^{-1}$) & SASA (nm$^2$)\\ 
 \hline\hline
 Helix    &  0.7555 $\pm$ 0.0007    & 0.2153 $\pm$ 0.0009          &  0.0338 $\pm$ 0.0004 & 9.8 $\pm$ 0.14\\  
 Hairpin  &  0.7474 $\pm$ 0.0007    & 0.2293 $\pm$ 0.0008          &  0.0329 $\pm$ 0.0004 & 9.8 $\pm$ 0.16\\
 Coil     &  0.7539 $\pm$ 0.0007    & 0.2209 $\pm$ 0.0009          &  0.0362 $\pm$ 0.0004 & 12.3 $\pm$ 0.15 \\
 \hline
\end{tabular}
\end{center}
\justify {\bf\todo{Supplementary Table 1.}} {\bf Cooling fit parameters.} Parameters for the cooling model $E(t) = E_\text{B} + [E(0) - E_\text{B}] e^{-\todo{k_\text{c}} t}$, measured with respect to the peak kinetic energy content $E(0)$ of Aib$_{10}$ immediately following excitation.  Data is provided for 750--member ensembles of helix, hairpin, or extended coil conformers at $\todo{T_\text{B}} = 230.0$ K, with the mean transport profile coarse--grained into 10 fs bins prior to analysis.  Fit error bands correspond to variation at the 95\% confidence interval.  The solvent--accessible surface area (SASA) calculated with a 0.1 nm probe is also provided for each conformer.  In this case, the bands correspond to plus/minus one standard deviation for the ensemble composed of the initial equilibrium conformations for the NEMD simulation.     
\end{table}

\clearpage 

\vspace{2\baselineskip}
\begin{figure}[h]
\bgroup
\setlength\tabcolsep{8.0pt}
\begin{tabular}{ccc}
\includegraphics[scale=0.95]{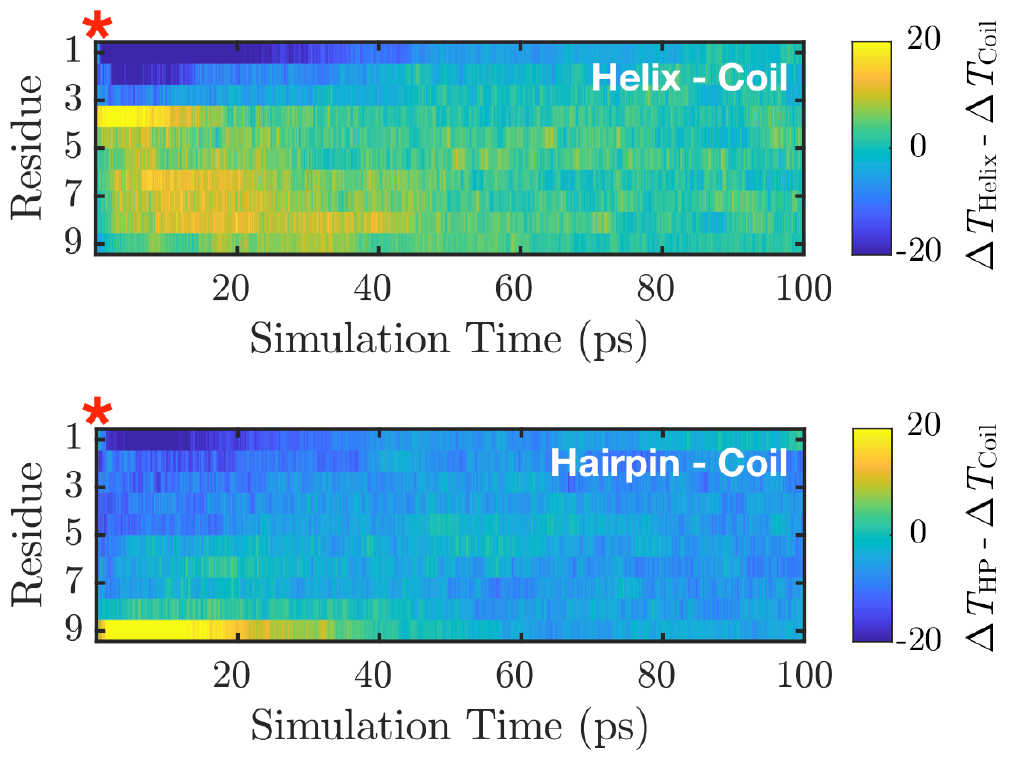}
\end{tabular}
\egroup
\justify {\bf\todo{Supplementary Figure 17.}} {\bf Variation between transport profiles.} Conformational dependence of the heat transport profile for helical, hairpin, and extended coil Aib$_{10}$ ensembles during cooling simulations at $\todo{T_\text{B}} = 230.0$ K.  Raw data are quantified through the ensemble--averaged temperature elevation $\langle T_j \rangle$ over the bath temperature $\todo{\Delta T_{\text{B},j}} = \langle T_{j}\rangle - \todo{T_\text{B}}$ at residue $j$ for designated conformational populations (following Fig.~1 of the main text).  Plots correspond to a difference map of these elevations between helical and coiled ($\Delta T_{\text{Helix},j} - \Delta T_{\text{Coil},j}$), as well as hairpin and coiled ($\Delta T_{\text{HP},j} - \Delta T_{\text{Coil},j}$) structures.  Heat redistribution throughout the peptide is more efficacious within helical and hairpin ensembles. For plotting purposes, the upper and lower temperature elevation (e.g., $\todo{\Delta T_{\text{B},j}}$) bounds are a cutoff for all values lying outside the range.  The heater site is denoted by a red asterisk.  
\label{fig:transProfile}
\end{figure}

\clearpage